\documentclass[aps,prd,twocolumn,groupedaddress,nofootinbib,preprintnumbers,superscriptaddress]{revtex4-2}  
\usepackage{graphicx}
\usepackage{amsmath,amssymb,latexsym,bbold}
\usepackage{amsfonts}
\usepackage[colorlinks=true,citecolor=red,linkcolor=blue,breaklinks=true]{hyperref}
\usepackage[figure, table]{hypcap}
\usepackage{xcolor}
\usepackage{subfigure}
\usepackage{setspace}
\usepackage{footnote}
\usepackage{multirow}
\usepackage[normalem]{ulem}
\usepackage[utf8]{inputenc}
\usepackage{mathrsfs}
\usepackage{slashed}
\usepackage{longtable}
\usepackage{enumitem}
\usepackage[scr=boondox]{mathalpha}
\usepackage[many]{tcolorbox}
\usepackage{rotating}
\usepackage{array}

\numberwithin{equation}{section}

\allowdisplaybreaks

\definecolor{tabred}{rgb}{0.8392156862745098, 0.15294117647058825, 0.1568627450980392}
\definecolor{tabblue}{rgb}{0.12156862745098039, 0.4666666666666667, 0.7058823529411765}
\hypersetup{colorlinks=true, linkcolor={tabred}, citecolor={tabblue}, urlcolor={tabblue}} 

\usepackage{orcidlink}

\usepackage{fontawesome5}
\definecolor{color_git}{rgb}{0.098, 0.160, 0.345}
\newcommand{\gitlink}{\href{https://github.com/mariofnavarro/COFLASY/tree/COFLASY-C}{\textsc{g}it\textsc{h}ub {\large\color{color_git}\faGithub}}}

\newcommand{\orcid}[1]{\,\orcidlink{#1}}

\newcommand{\dd}{\mathrm{d}}
\newcommand{\fd}{f_\mathrm{fd}}

\begin{document}

\preprint{CERN-TH-2025-192}
\preprint{LA-UR-25-29666}

\vspace*{1cm}
\title{A Limit on the Total Lepton Number in the Universe from BBN and the CMB}

\author{Valerie Domcke\orcid{0000-0002-7208-4464}}
\email{valerie.domcke@cern.ch}
\affiliation{Theoretical Physics Department, CERN, 1 Esplanade des Particules, CH-1211 Geneva 23, Switzerland}

\author{Miguel Escudero\orcid{0000-0002-4487-8742}}
\email{miguel.escudero@cern.ch}
\affiliation{Theoretical Physics Department, CERN, 1 Esplanade des Particules, CH-1211 Geneva 23, Switzerland}

\author{Mario Fern\'andez Navarro\orcid{0000-0002-8796-0172}}
\email{mario.fernandeznavarro@glasgow.ac.uk}
\affiliation{School of Physics \& Astronomy, University of Glasgow, Glasgow G12 8QQ, UK}

\author{Stefan Sandner\orcid{0000-0002-1802-9018}\,}
\email{stefan.sandner@lanl.gov}
\affiliation{Theoretical Division, Los Alamos National Laboratory, Los Alamos, NM 87545, USA}

\begin{abstract}
\noindent
At temperatures below the QCD phase transition, any substantial lepton number in the Universe can only be present within the neutrino sector. In this work, we systematically explore the impact of a non-vanishing lepton number on Big Bang Nucleosynthesis (BBN) and the Cosmic Microwave Background (CMB). Relying on our recently developed framework based on momentum averaged quantum kinetic equations for the neutrino density matrix, we solve the full BBN reaction network to obtain the abundances of primordial elements. We find that the maximal primordial total lepton number $L$ allowed by BBN and the CMB is $-0.12 \,(-0.10) \leq  L \leq 0.13\,(0.12) $ for NH (IH), while specific flavor directions can be even more constrained. This bound is complementary to the limits obtained from avoiding baryon overproduction through sphaleron processes at the electroweak phase transition since, although numerically weaker, it applies at lower temperatures and is obtained completely independently. We publicly release the C++ code \texttt{COFLASY-C} on~\gitlink $\,$ which solves for the evolution of the neutrino quantum kinetic equations numerically.
\end{abstract}

\maketitle

\newpage
{
\hypersetup{linkcolor=black}
\tableofcontents
}

\vspace{1 mm}
\vspace{-4.5mm}

\setlength{\parskip}{1pt}

\section{Introduction}

The matter antimatter asymmetry observed in our Universe and crucial to our existence is clear evidence of $CP$ violation playing a key role in the evolution of the Universe. While the observed baryon asymmetry today is very small, $\eta_B \equiv n_B/n_\gamma\simeq 6\times 10^{-10}$~\cite{Planck:2018vyg}, there are some hints for large $CP$ violation in our cosmic history, in the form of helical magnetic fields proposed to explain the lack of secondary photons in blazar observations~\cite{Dermer:2010mm,Taylor:2011bn,MAGIC:2022piy}, possible birefringence in the Cosmic Microwave Background (CMB)~\cite{Minami:2020odp}, or a sizable asymmetry in electron neutrinos which would address anomalously low primordial helium abundance measurements~\cite{Matsumoto:2022tlr,Yanagisawa:2025mgx,Burns:2022hkq,Escudero:2022okz,March-Russell:1999hpw}. Moreover, several baryogenesis models invoke large primordial asymmetries in order to explain the small relic asymmetry observed today~\cite{Cohen:1987vi,Cohen:1988kt,Davidson:1994gn,March-Russell:1999hpw,Yamaguchi:2002vw,Kawasaki:2002hq,Chiba:2003vp,Takahashi:2003db,Asaka:2005pn,Shaposhnikov:2008pf,Laine:2008pg,Kamada:2018tcs,Domcke:2019mnd,Domcke:2020quw,Mukaida:2021sgv}. This motivates the following question: What is the maximal lepton number generated in the early Universe that is compatible with Big Bang Nucleosynthesis (BBN) and CMB observations? To our knowledge, this is the first study to systemically address this question.

Not surprisingly, the bound we will find is numerically much weaker than the bound on lepton number obtained at the electroweak scale, when sphaleron processes convert any difference of baryon and lepton number to a baryon asymmetry. However, we stress that the limits derived here are fully independent, and apply at a lower temperature. They hence apply also in scenarios in which an asymmetry is generated after the electroweak phase transition~\cite{Yamaguchi:2002vw,Shaposhnikov:2008pf,Akita:2025zvq} or in which the Universe has a very low reheating temperature (but above $20\,\mathrm{MeV}$).

At temperatures below the QCD phase transition, any substantial lepton number in the Universe can only be present within the neutrino sector. Neutrino oscillations tend to equilibrate the lepton asymmetries stored in each flavor. Neutrino oscillations start at $T\sim 15\,\mathrm{MeV}$ and are efficient in equilibrating the three neutrino flavors down to $T\sim 2\,{\rm MeV}$ when neutrinos decouple. After this point, any residual asymmetry in electron neutrinos will strongly impact the neutron-to-proton ratio, which then in turn alters the nuclear reaction network of BBN, which is the seed for the abundances of all primordial elements~\cite{Dimopoulos:1979ma,Schramm:1979kn,Bernstein:1988ad,Pitrou:2018cgg}. Critically, due to the neutron-proton weak interactions that are active until $T_\gamma \simeq 0.7\,{\rm MeV}$, an excess of electron neutrinos over electron anti-neutrinos will lead to a smaller number density of neutrons in the Universe as compared with the Standard Model prediction and hence to a smaller helium abundance in the Universe. Moreover, the flavor equilibration process leads to entropy injection and therefore to a relative increase of the neutrino temperature, which is partially transmitted to the photon bath as long as the neutrinos are not yet fully decoupled from the plasma. Relating the resulting light element abundances as well as the total energy in the neutrino sector (parametrized by the effective number of neutrinos $N_{\rm eff}$) to the asymmetries in the neutrino sector at $T\sim 15\,\mathrm{MeV}$ thus requires a detailed understanding of the evolution of the neutrino distributions in this time window.

We tackle this by solving the quantum kinetic equations governing the evolution of the neutrino density matrices~\cite{Sigl:1993ctk} with a momentum-averaged ansatz~\cite{Domcke:2025lzg}. As demonstrated in~\cite{Domcke:2025lzg} and verified by cross-checking against the momentum dependent code used in Refs.~\cite{Froustey:2021azz, Froustey:2020mcq,Froustey:2024mgf}, this method is both efficient and reliable, allowing for a systematic coverage of the parameter space. Solving the Boltzmann equations of the BBN reaction network, we then map the asymptotic electron neutrino asymmetry, the neutrino temperature and the photon temperature to predict the light element abundances. Combining the observed helium abundance with constraints on $\Delta N_\text{eff}$ breaks degeneracies in the flavor parameter space and results in a limit $-0.12 \,(-0.10) \leq  L \leq 0.13\,(0.12) $ for normal (inverted) hierarchy of the neutrino masses, with lepton number $L$ defined in Eq.~\eqref{eq:L_definition}. As a byproduct of our analysis, we show that neutrino lepton number flavor asymmetries in the Universe today can be as large as $\sim 35 /{\rm cm}^{3}$, which may be relevant for some direct detection techniques of the Cosmic Neutrino Background. Last, but not least, we highlight that we release the C++ code \texttt{COFLASY-C} available on \gitlink\, which implements and efficiently solves the quantum kinetic equations numerically.

\section{Framework}

In order to asses the cosmological impact of primordial lepton asymmetries we need to address how the latter are redistributed and damped by neutrino oscillations in the temperature range from $T\sim 15\,{\rm MeV} $ to $T\sim 2\,{\rm MeV}$ and how the surviving residual lepton asymmetries impact the number of relativisitic neutrino species $N_{\rm  eff}$ and the neutron-to-proton ratio in the Universe. To this end, we will use 1) the momentum-averaged approach developed and validated in~\cite{Domcke:2025lzg} to track the neutrino ensemble evolution in the presence of large primordial asymmetries, and 2) study their subsequent impact on BBN by tracking all relevant primordial element abundances. We elaborate on our methodology in this section.

\subsection{Neutrino Quantum Kinetic Equations}\label{sec:nuQKE}

The time evolution of a neutrino ensemble in a thermal environment in the presence of neutrino oscillations needs to account for quantum correlations between different flavor states on top of tracking their classical phase space densities. This is most conveniently done via the quantum kinetic equations (QKEs) for the momentum ($k = |\vec{k}|$) and time ($t$) dependent density matrices $\rho$ ($\bar \rho$) of (anti-)neutrinos~\cite{Sigl:1993ctk,Volpe:2013uxl,Blaschke:2016xxt,Bennett:2020zkv,Froustey:2021azz, Froustey:2020mcq, Li:2024gzf}:
\begin{align}
\label{eq:QKE}
\frac{d \rho}{dt} = - i [{\cal H}, \rho] + \mathcal{I}\left[ \rho\right] \,, \;\,  \frac{d \bar \rho}{dt} = +i [{\cal H}, \bar  \rho]  + \bar{\mathcal{I}}\left[ \bar{\rho}\right] \,,
\end{align}
which relate to the physical neutrino number and energy densities respectively
\begin{align}
\!\!\!\!    \overset{(-)}{n}_\alpha =\int \frac{\dd^3k}{(2\pi)^3} \overset{(-)}{\rho}_{\alpha\alpha}\,, \;\;  \varepsilon_\nu = \int \frac{\mathrm{d}^3 k }{(2\pi)^3}  \, k \,{\rm Tr}[ \rho+\bar{\rho}]\,.
\end{align}
The Hamiltonian $\mathcal{H} = {\cal H}_0 + {\cal V}$ contains the vacuum evolution operator 
\begin{align}
    {\cal H}_0 = U \frac{M^2}{2k} U^\dagger \,,
\end{align}
where $M$ is the diagonal light neutrino mass matrix and $U$ the PMNS matrix~\cite{ParticleDataGroup:2024cfk}.\footnote{In the PMNS matrix we fix all mixing angles to their best fit values and assume no CP violation. The numerical values are given in~\cite{Domcke:2025lzg}. We remark that uncertainties in the mixing angles have a very minor impact on the final asymmetries, with the possible exception of $\theta_{13}$, and they are also largely insensitive to the CP-violating phase.} The matter potentials, ${\cal V} = V_c \pm V_s$, with $+(-)$ for (anti-)neutrinos, are flavor $\alpha = \{e,\mu,\tau\}$ dependent and account for neutrino self-energy corrections arising from charged and neutral leptons respectively~\cite{Sigl:1993ctk,Notzold:1987ik,Froustey:2020mcq}:
\begin{align}
    V_{\mathrm{c}} &=  - 2 \sqrt{2} G_F \frac{k}{m_W^2}  (\varepsilon_\alpha + p_\alpha) \,,\\
    V_{\mathrm{s}} & = \sqrt{2} G_F  \int \frac{\mathrm{d}^3k'}{(2 \pi)^3} (\rho - \bar{\rho}) \,,
\end{align}
where $V_\mathrm{c}$ contains the summed energy and pressure densities of both charged leptons and anti-leptons~\cite{Domcke:2025lzg}. The collision terms $\mathcal{I}\,(\bar{\mathcal{I}})$ and the Hamiltonian conserve total lepton number which we define as:
\begin{align}\label{eq:L_definition}
    L \equiv \sum_\alpha \frac{n_\alpha - \bar{n}_\alpha}{T_{\rm cm}^3} = \mathrm{const.} \,,
\end{align}
i.e.  $\sum_\alpha (n_\alpha-\bar{n}_\alpha) x^3 = {\rm constant}$, where we introduced the scale factor $x$ as a measure of time or comoving temperature $T_{\mathrm{cm}}$
\begin{align}
    x = T_{\mathrm{ref}}/T_{\mathrm{cm}}\,,
\end{align}
and $T_\mathrm{ref}$ is an arbitrary reference temperature. The lepton number $L$ is related to the lepton-to-entropy density ratio in the Universe
\begin{align}
    Y_L \equiv \sum_\alpha \frac{n_\alpha - \bar{n}_\alpha}{s} \simeq \frac{1}{4.7} \,L\,,
\end{align}
where we note that since large lepton asymmetry equilibration leads to entropy injection this quantity can decrease by up to $5\%$ thorough the course of neutrino oscillations in the parameter space of interest.

The collision terms $\mathcal{I}\,(\bar{\mathcal{I}})$ incorporate $\nu e \leftrightarrow \nu e$ scatterings, $\nu \bar{\nu} \leftrightarrow e^+e^-$ annihilation and neutrino self-interactions $\nu\nu \leftrightarrow \nu \nu$ and $\nu\bar{\nu} \leftrightarrow \nu \bar{\nu}$. The collision terms involve double integrals which depend on $k$ and are non-linear in $\rho$~\cite{Sigl:1993ctk,Volpe:2013uxl,Blaschke:2016xxt,Bennett:2020zkv,Froustey:2021azz, Froustey:2020mcq, Li:2024gzf}. This, together with the simultaneous presence of vastly different timescales in $\mathcal{H}$, makes it challenging to numerically solve the system~\cite{Bennett:2020zkv,Akita:2020szl,Froustey:2021azz,Froustey:2024mgf,deSalas:2016ztq, Li:2024gzf}. However, as we demonstrated in~\cite{Domcke:2025lzg}, momentum-averaged QKE are a good approximation to the full system, while reducing the computational cost by at least a factor of $\mathcal{O}(10^3)$.
We therefore parametrize 
\begin{align}
\label{eq:ansatz}
 \rho(x,y) = r(x) \, f_\mathrm{fd}(y/z_\nu) \,,
\end{align}
with $f_\mathrm{fd}$ the Fermi-Dirac distribution without chemical potential and the dimensionless variables
\begin{align}\label{eq:definitions}
y = k/T_\mathrm{cm}\,,\;\; z_i = T_i / T_{\mathrm{cm}}\,,
\end{align}
with $i =\{\nu,\gamma\}$. For $\bar{\rho}$ we have the same but with $r\mapsto \bar{r}$.

The ansatz of Eq.~\eqref{eq:ansatz} allows to explicitly integrate over $y$, leading to the momentum averaged QKE
\begin{align}
\label{eq:kinetic_eq_flavor}
\begin{split}
     Hx r' + 3 H x \, r \, \frac{z_\nu'}{z_\nu}  &= -i [\langle \mathcal{H} \rangle  , r] + \langle \mathcal{I} \rangle\,,
\end{split}
\end{align}
and similar for $\bar{r}$, with prime denoting derivatives with respect to $x$. The momentum average corresponds to 
\begin{align}
\label{SM:eq:def_momentum_ave}
    \langle ... \rangle \equiv  \frac{2}{3 \zeta(3) z_\nu^3} \int\mathrm{d}y\, ... y^2 \fd(y/z_\nu) \,,
\end{align}
with $\zeta(3) \simeq 1.20206$. The Hubble parameter is
\begin{align}
\label{eq:Hubble}
    H^2 = \frac{\varepsilon_\gamma + \varepsilon_e + \varepsilon_\nu}{3 M_P^2}\,,
\end{align}
with the reduced Planck mass $M_P = 2.435 \cdot 10^{18}\,\mathrm{GeV}$. In addition, we need to track the energy transfer between the neutrino sector and electron-photon plasma. We do this via the continuity equation which relates the change in energy is both sectors,
\begin{align}
\label{eq:z_evolution}
\begin{split}
 - \frac{\delta \varepsilon}{\delta t} &= \frac{d (\varepsilon_\gamma + \varepsilon_e)}{dt} + 4 H \varepsilon_\gamma + 3 H (\varepsilon_e + p_e) \,, \\
 \frac{\delta \varepsilon}{\delta t} &= \frac{d \varepsilon_\nu}{dt} + 4 H \varepsilon_\nu \,.
\end{split}
\end{align}
The energy transfer, $\delta \varepsilon/ \delta t$, is determined by the diagonal elements of the collision terms,
\begin{align}
\label{eq:def_energy_transfer}
 \frac{\delta \varepsilon}{\delta t} =  \int \frac{\dd^3 k}{(2\pi)^3} \, k   \, \text{Tr}[ {\cal I} + \bar{\cal I}] \,.
\end{align}
The explicit form of the momentum averaged collision terms and the energy transfer is not particularly compact nor illuminating and we refer to~\cite{Domcke:2025lzg} for a detailed derivation. The physical initial conditions are set at $T = 20\,\mathrm{MeV}$ and correspond to an \textit{overdamped} neutrino state which is tightly coupled to the electron-photon plasma, such that
\begin{align}
T_\mathrm{cm}^\mathrm{ini} = T_\gamma^\mathrm{ini} = T_\nu^\mathrm{ini} \,,\; 
    r_{\alpha \beta}^\mathrm{ini} = \bar{r}_{\alpha \beta}^\mathrm{ini} = 0\, \,\text{with} \,\alpha \neq \beta\,\,\,.
\end{align}
On the other hand, the presence of a primordial lepton asymmetry changes the (anti-)neutrino densities which is captured by the diagonal elements of $r$. For our momentum averaged ansatz this implies 
\begin{align}
\label{eq:ini_flavor_diag}
r_{\alpha \alpha}^\mathrm{ini} = - \frac{4 }{3 \zeta(3)} \mathrm{Li}_{3}\left( -\mathrm{e}^{\xi_\alpha}\right)
\,,
\end{align}
with $\xi_\alpha = \mu_\alpha/T_\mathrm{cm}$ and for $\bar{r}$ we have $\xi_\alpha \mapsto -\xi_\alpha$. $\mathrm{Li}_{3}$ is a polylogarithm and for a more detailed discussion on the initial conditions for momentum averaged QKE we refer to~\cite{Domcke:2025lzg}. This corresponds to a total lepton number of $L = \sum_\alpha  (\xi_\alpha / 6+\xi_\alpha^3/(6\pi^2))$.

\begin{figure}[!t]
\includegraphics[width=0.475\textwidth]{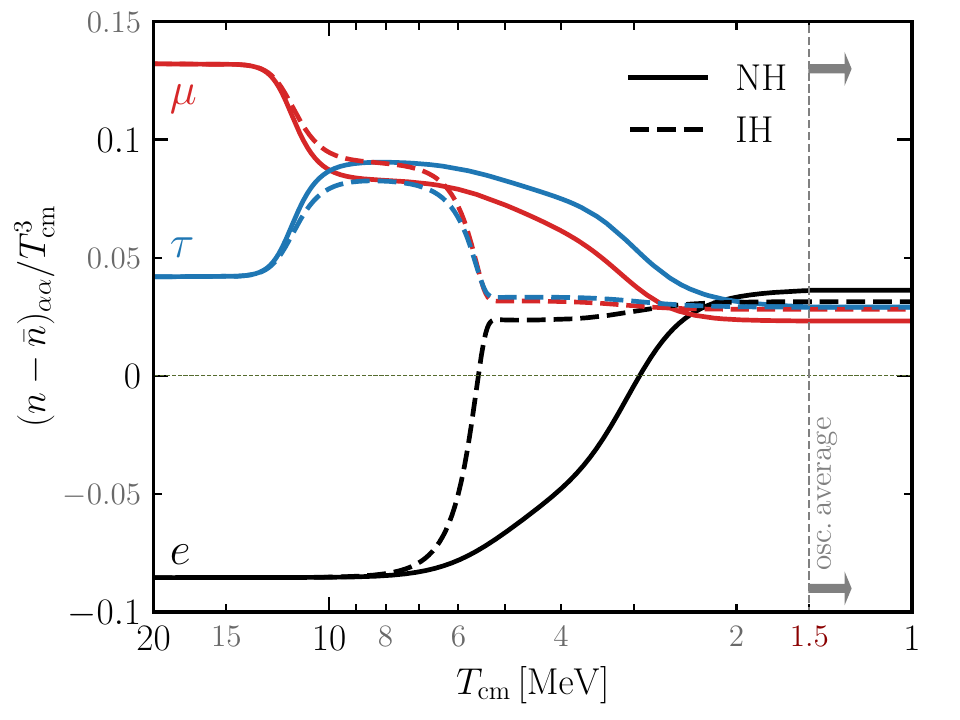}
\caption{Typical example of the time evolution for large amplitude of initial flavor asymmetries ($\xi_e= -0.5, \xi_\mu= 0.75, \xi_\tau=0.25$) in NH (solid) and IH (dashed).
The full solution depicted here essentially overlaps with the adiabatic approximation $(V_s =0)$~\cite{Domcke:2025lzg}.
}
\label{fig:time-evolution}
\end{figure}

Solving Eq.~\eqref{eq:kinetic_eq_flavor} together with Eq.~\eqref{eq:z_evolution} allows us to extract the relevant quantities which alter the neutron-to-proton ratio, i.e. the (anti-)neutrino number and energy density
\begin{align}
    \overset{(-)\;\;}{n_\alpha} &= \frac{3 \zeta(3)}{4\pi^2} T_\nu^3\, \overset{(-)\;\;\;\;}{r_{\alpha\alpha}} \,, \\
    N_\mathrm{eff} &= \left( \frac{11}{4} \right)^{4/3}  \left(\frac{z_\nu}{z_\gamma} \right)^4 \frac{\mathrm{Tr}[r + \bar r]}{2}\,,
\end{align}
as well as the Hubble parameter of Eq.~\eqref{eq:Hubble}. Here $N_{\mathrm{eff}}$ is defined as relevant for CMB observations, namely, at $T_\gamma \ll m_e$, and the excess radiation is parametrized by $\Delta N_{\mathrm{eff}} = N_{\mathrm{eff}} - N_{\mathrm{eff}}^{\mathrm{SM}}$ with $N_{\mathrm{eff}}^{\mathrm{SM}} = 3.044$ the SM value~\cite{Akita:2020szl,Froustey:2020mcq,Bennett:2020zkv,EscuderoAbenza:2020cmq,Jackson:2024gtr,Ihnatenko:2025kew,Escudero:2025mvt}.
An exemplary evolution of the neutrino number density asymmetry is shown in Fig.~\ref{fig:time-evolution}.
It is intended to depict a \textit{typical} time evolution, but we remark that the QKEs lead to much richer structure in specific flavor directions, which have been discussed in detail in~\cite{Domcke:2025lzg}. See also App.~\ref{sec:nonadiabatic} in which we uncover a previously overlooked phenomena of non-adiabatic oscillations at late times. We further refer to \gitlink\, where we make available our numerical code solving the QKEs.

Having set up the stage to solve for the neutrino evolution, we continue in the following subsection to discuss its impact on the neutron-to-proton ratio and ultimately on the formation of the light element abundances.

\subsection{Big Bang Nucleosynthesis Analysis}
\label{subsec:BBN_Analysis}

\begin{figure}[!t]
\includegraphics[width=0.475\textwidth]{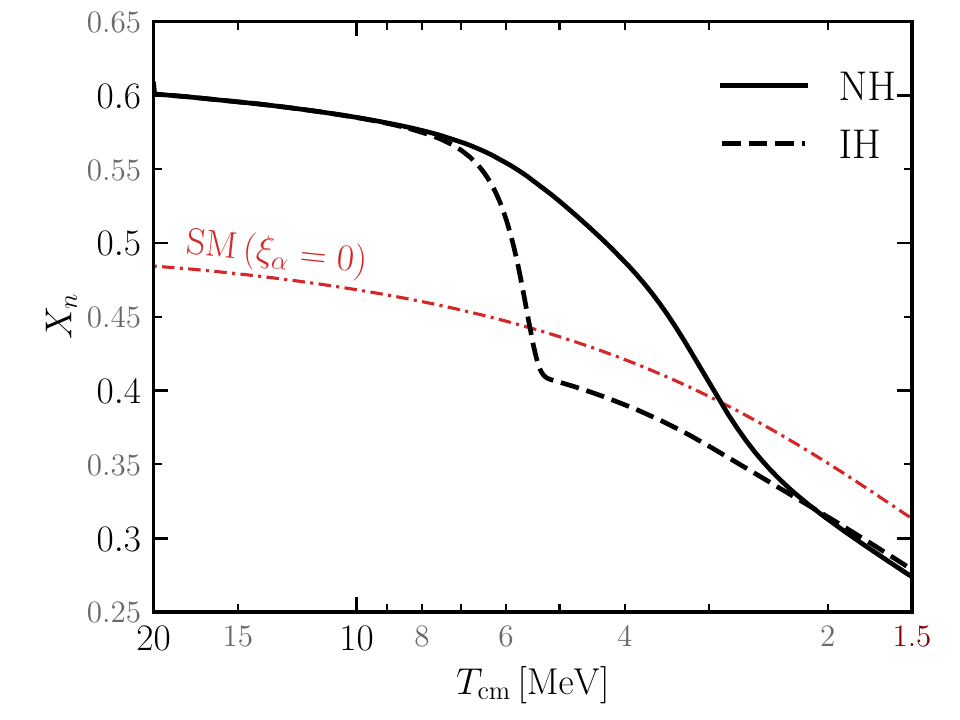}
\caption{Impact of neutrino asymmetries on the evolution of the neutron fraction in the plasma for the same initial conditions as in Fig.~\ref{fig:time-evolution}. }
\label{fig:Xn-example}
\end{figure}

\begin{figure}[!t]
\includegraphics[width=0.475\textwidth]{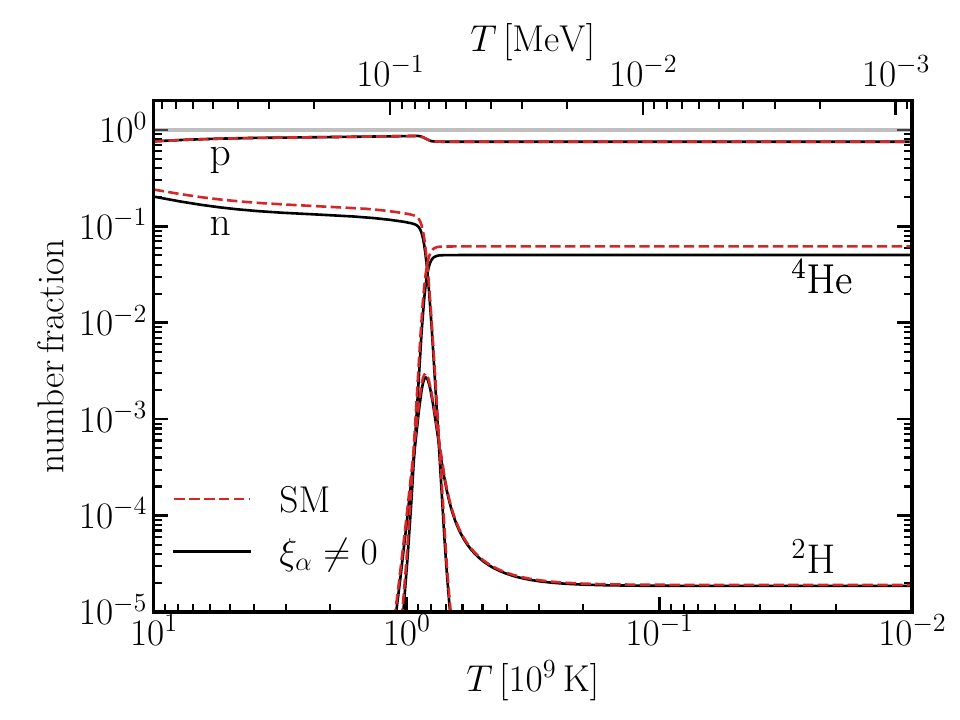}
\caption{
Prediction of the primordial element abundances within the SM (red dashed) compared to a scenario with non-vanishing neutrino chemical potential (black solid). Specifically we choose NH and the same initial conditions are in Fig.~\ref{fig:time-evolution} and Fig.~\ref{fig:Xn-example}. The IH result is very similar. 
We use $\eta_B = 6.109\times 10^{-10}$ and $\tau_n = 878.4\,{\rm s}$ for the BBN network. Note that the helium mass fraction is $4\times $ the number fraction shown here. 
}
\label{fig:BBN-example}
\end{figure}

\textit{Light Element Abundances --}
Neutrinos and anti-neutrinos directly participate in the weak processes that interconvert protons and neutrons in the early Universe. The relevant interactions are $n + \nu_e \leftrightarrow p + e^-$, $n +e^+ \leftrightarrow p + \bar{\nu}_e$ and $n \leftrightarrow p + e^- + \bar{\nu}_e$ and as such an excess of $\nu_e$ over $\bar{\nu}_e$ would lead to a reduced number of neutrons with respect to the standard BBN prediction. At temperatures $T\gtrsim 0.2\,{\rm MeV}$, the fractional number of neutrons in the plasma ($X_n = n_n/(n_n + n_p)$) is governed by the following equation~\cite{Bernstein:1988ad}:
\begin{align}
\label{eq:Xn_evol}
    \frac{d X_n}{dt} = - \lambda_{np} X_n + \lambda_{pn} (1-X_n)\,,
\end{align}
where $\lambda_{np}$ and $\lambda_{pn}$ are the rates for neutron-to-proton and proton-to-neutron interconversion, respectively. We give their explicit form in App.~\ref{sec:np_pn_rates}, noting in particular that they depend on the distribution functions of $\nu_e$ and $\bar{\nu}_e$ neutrinos. As demonstrated in~\cite{Domcke:2025lzg}, the momentum averaged QKEs track the physical quantities of the (anti-)neutrino system, i.e.\ their number and energy densities, to high accuracy. In order to describe the neutron-proton interactions we assume neutrinos follow a Fermi-Dirac distribution with an effective chemical potential $\xi_e$ and a temperature $T_{\nu_e}$, which we extract from the number and energy densities obtained from our QKEs solver.
We find this procedure to be more accurate in modeling the neutron-proton weak interactions than using the ansatz~\eqref{eq:ansatz} in our BBN analysis. 

Concretely, our approach is to solve the QKE for the neutrinos using the ansatz~\eqref{eq:ansatz} and map the results for the number and energy densities of neutrinos and anti-neutrinos into an effective chemical potential and temperature for $\nu_e$-$\bar{\nu}_e$. Together with $N_{\rm eff}$ we then calculate the primordial element abundances. We expect this to be a very good description to the actual microphysical state of electron neutrinos since the starting point of the QKE are purely thermal distribution functions and neutrino oscillations tend to equilibrate well before neutrinos decouple at $T\simeq 2\,{\rm MeV}$. Critically, it ensures that the number and energy densities match the outcome of the QKEs, while respecting detailed balance in the neutron-proton interactions as relevant for BBN.
Working at leading order in the small chemical potentials one finds:
\begin{subequations}\label{eq:MapEqs}
    \begin{align}
    T_{\nu_e}/T_\gamma &= \left(\frac{\frac{\rho_{\nu_e} + \rho_{{\bar \nu}_e}}{T_{\rm cm}^4}}{2\frac{7}{8} \frac{\pi^2}{30}}\right)^{1/4} \frac{1}{z_\gamma}\,, \label{eq:Tnumap}\\
    \xi_e &= 6\times \frac{n_{\nu_e} - n_{{\bar \nu}_e}}{T_{\rm cm}^3} \times \left(\frac{\frac{\rho_{\nu_e} + \rho_{{\bar \nu}_e}}{T_{\rm cm}^4}}{2\frac{7}{8} \frac{\pi^2}{30}}\right)^{-3/4} \label{eq:xiemap}\,,
\end{align}
\end{subequations}
where $({\rho_{\nu_e} + \rho_{{\bar \nu}_e}})/{T_{\rm cm}^4}$, $({n_{\nu_e} - n_{{\bar \nu}_e}})/{T_{\rm cm}^3}$ and $z_\gamma$ are direct outputs of our solver of the QKEs. We explicitly checked that next-to-leading order $\xi_e$ contributions do not alter any of our results. We note that the parameters in Eq.~\eqref{eq:MapEqs} are effective parameters in that 1) at the relevant temperatures for BBN, neutrinos are actually propagating in the mass basis, and 2) we approximate their momentum distribution by a Fermi-Dirac function.

Neutrinos are safely decoupled by $T\simeq 1.5\,{\rm MeV}$, while the proton-neutron interactions freeze-out much later at $T\simeq 0.7\,{\rm MeV}$. Thus for $T < 1.5\,{\rm MeV}$, $T_{\nu_e}$ and $\xi_e$ evolve only adiabatically and we have $d\xi_e/dt = 0$ and $dT_{\nu_e}/dt = - H \, T_{\nu_e}$. This means that in practice we can solve for the neutron fraction in the plasma $X_n$ \textit{after} we have found the solution of the residual lepton asymmetries, i.e. Eq.~\eqref{eq:QKE} and Eq.~\eqref{eq:Xn_evol} are treated as decoupled. In Fig.~\ref{fig:Xn-example} we show the evolution of the neutron number density using the same neutrino flavor evolution as in Fig.~\ref{fig:time-evolution}. From these figures we can appreciate that the (non-trivial) neutrino flavor evolution effectively stops at $T\simeq (2-3)\,{\rm MeV}$. From there onward the neutrino evolution is adiabatic and hence this provides a good description of the neutron number density in the plasma as neutron-proton interactions freeze-out significantly later at $T\simeq 0.7\,{\rm MeV}$.

Radiative corrections and the effect of a finite nucleon mass in $p\leftrightarrow n$ reactions are needed in order to predict with high accuracy the primordial element abundances~\cite{Lopez:1997ki,Iocco:2008va,Pitrou:2018cgg}. These effects are non-trivial to include in the calculation but are not expected to depend appreciatively on a possible non-standard neutrino distribution function. In consequence and for simplicity, we in practice solve $X_n$ using the Born rates without including these small effects, but then we take them into account as a universal correction factor for the rates as calculated in \texttt{PRIMAT}~\cite{Pitrou:2019nub,Pitrou:2018cgg} assuming the Standard Model evolution, see App.~\ref{sec:np_pn_rates}. To calculate the subsequent evolution and the eventual formation of ${\rm D}$, ${\rm T}$, $^3{}{\rm He}$, $^4{}{\rm He}$, $^7{}{\rm Li}$ we use our own BBN code written in \texttt{Mathematica} and which is greatly inspired by the \texttt{PRIMAT}~\cite{Pitrou:2019nub,Pitrou:2018cgg} and \texttt{PRyMordial}~\cite{Burns:2023sgx} BBN codes. We account for the small nuclear reaction network of $12$ nuclear reactions and use the tabulated rates as a function of $T_\gamma$ taken from \texttt{PRIMAT}. For the Standard Model, our results agree very well to those of \texttt{PRIMAT} but also to those of \texttt{PArthENoPE}~\cite{Pisanti:2007hk,Consiglio:2017pot,Gariazzo:2021iiu} when using their nuclear reaction rates for ${\rm D}+{\rm D}\to {}^3 {\rm He} + n$ and ${\rm D}+{\rm D}\to {\rm T} + p$ as reported in the online repository of \texttt{PRIMAT}. In particular, when solving our BBN reaction network in the Standard Model we agree with the results of the \texttt{PRIMAT} code with an accuracy of $0.2\%$ for $Y_{\rm p}$ and $0.4\%$ for deuterium which are negligible errors compared with either observational uncertainties or those arising from nuclear reaction rates. We show in Fig.~\ref{fig:BBN-example} a numerical example highlighting the mass fractions of the lightest elements using the same initial conditions as in Figs.~\ref{fig:time-evolution} and~\ref{fig:Xn-example}.

Following this approach, we are able to solve for the primordial element abundances and obtain convenient fitting functions depending on the cosmological parameters we are interested in:
\begin{widetext}
\begin{subequations}
\label{eq:fittingfuncs}
    \begin{align}
    \begin{split}\label{eq:fittingfuncs_YP}
\!\!\!\!\!\! \!\!\!\!\!\!        \frac{Y_{\rm P}}{Y_{\rm P}^{\rm Ref}}  =  \left[\frac{\Omega_b h^2}{\Omega_b h^2|^{\rm Ref}}\right]^{0.039}  \exp \bigg\{ &-0.961\, \xi_e -0.185\, \xi_e^2 +0.001\, \xi_e^3 + 0.053\,\Delta N_{\rm eff}  -0.682\, \delta_T+0.72\,\delta_T^2  \\ 
        &+  0.011 \,\xi_e \Delta N_{\rm eff} -3.5 \,\xi_e \delta_T-0.9\,\xi_e^2 \delta_T+2.2\,\xi_e \delta_T^2+ 0.016\, \Delta N_{\rm eff} \delta_T \bigg\} \,, 
    \end{split}
        \\
    \begin{split}\label{eq:fittingfuncs_DH}
\frac{{\rm D/H}|_{\rm P}}{\,\,{\rm D/H}|_{\rm P}^{\rm Ref}} =  \left[\frac{\Omega_b h^2}{\Omega_b h^2|^{\rm Ref}}\right]^{-1.64}\exp \bigg\{ & -0.528\, \xi_e + 0.22 \,\xi_e^2 -0.052\, \xi_e^3 + 0.133\,\Delta N_{\rm eff}  -0.378 \,\delta_T+0.562 \,\delta_T^2  \\ 
        &- 0.026\, \xi_e \Delta N_{\rm eff} -1.5 \,\xi_e \delta_T- 0.0096 \,\Delta N_{\rm eff} \delta_T  \bigg\}  \,.
    \end{split}
    \end{align}
\end{subequations}
\end{widetext}
In these expressions, $\xi_e$ is the asymptotic effective $\nu_e$ chemical potential as given by Eq.~\eqref{eq:xiemap}, and $\delta_T $ is the electron neutrino temperature ratio deviation with respect to the Standard Model expectation at $T\ll m_e$ i.e., $\delta_T \equiv {T_{\nu_e}}/{T_\gamma}-(11/4)^{-1/3}$. Note that the range we used to obtain the fitting functions is $|\xi_{e}|\leq0.4,\,|\delta_{T}|\leq0.071$ and $0\leq\Delta N_{\mathrm{eff}}\leq0.6$. To facilitate comparisons, we fix $Y_{\rm P}^{\rm Ref} = 0.2470$ and $10^5 \,{\rm D/H}|^{\rm Ref}_{\rm P} = 2.437$, as obtained from the \texttt{PRIMAT} code. Both $Y_{\rm P}$ and ${\rm D/H}|_{\rm P}$ depend upon $\Omega_b h^2$, and we will use the value of $\Omega_b h^2 = 0.02242 \pm 0.00014$ from Planck CMB observations. We note that the leading linear dependence on $\xi_e$ is the same as that found in~\cite{Froustey:2024mgf}, and that the powers for $\Omega_b h^2$ are in very good agreement with those output in~\cite{Fields:2019pfx}. The fitting formulas~\eqref{eq:fittingfuncs} are a stand-alone result of this work, applicable beyond our QKE solver as long as the definitions of $\xi_e$ and $\delta_T$, see Eq.~\eqref{eq:MapEqs}, are respected.

\vspace{2mm}
\textit{Validation of the Framework --} Our main assumption in solving the primordial element abundances is that the weak interactions that interconvert neutrons and protons are governed by a neutrinos following a Fermi-Dirac distribution function whose parameters are determined by the asymmetry in the neutrino number densities and the energy of the neutrino bath at $T < 1.5\,\mathrm{MeV}$. We have tested this assumption by comparing the output of our fitting functions in~\eqref{eq:fittingfuncs} with the full momentum dependent solutions as calculated in~\cite{Froustey:2024mgf} using the \texttt{PRIMAT} BBN code. Our results are summarized in App.~\ref{sec:BBNvalidation} and we see a good agreement for the whole parameter space except for some rather special regions in which the neutrino flavor evolution is non-adiabatic near $T\lesssim 3\,{\rm MeV}$. These regions only appear for specific flavor directions and large primordial lepton asymmetries. In consequence those regions are typically disfavored from $N_{\rm eff}$ CMB measurements, which means that they do not pose a problem for our analysis.

\vspace{2mm}
\textit{Cosmological Data Analysis -- } Having validated our framework we use the following data as a baseline to constrain the primordial lepton asymmetries: 
\begin{enumerate}
    \item Measurements of the primordial helium abundance. We use the recommended PDG determination of the primordial helium mass fraction~\cite{ParticleDataGroup:2024cfk}:
    \begin{align}\label{eq:Ypmeasurement}
        Y_{\rm P} \equiv \frac{4n_{\rm ^{4}He}}{n_{\rm H} + 4n_{\rm ^{4}He}} = 0.245 \pm 0.003\,.
    \end{align}
    \item The measurement of $N_{\rm eff}$ from CMB observations by Planck which reads~\cite{Planck:2018vyg}:
    \begin{align}
    \label{eq:NeffPlanck}
        N_{\rm eff} = 2.99\pm0.17\,.
    \end{align}
\end{enumerate}

As discussed in detail in App.~\ref{sec:BBNvalidation} for $Y_{\rm P}$ and in Ref.~\cite{Domcke:2025lzg} for $N_{\rm eff}$, our theoretical prediction uncertainty is small and in particular substantially smaller than the observational errors across the most relevant regions of parameter space. Therefore, and for simplicity, we neglect it in what follows.

We note that in principle we could also use the deuterium abundance to perform cosmological inferences but we do not implement it for three reasons: 1) deuterium contains only one neutron and therefore its abundance is substantially less sensitive to the primordial electron neutrino asymmetry than that of helium-4 (compare the exponents in Eq.~\eqref{eq:fittingfuncs_YP} and~\eqref{eq:fittingfuncs_DH}), 2) the deuterium abundance depends strongly upon the baryon density in the Universe, and 3) its theoretical prediction currently has a $\simeq 2-3\%$ uncertainty arising from nuclear reaction rate uncertainties~\cite{Pisanti:2020efz,Pitrou:2020etk,Yeh:2022heq}. Therefore, the role of the deuterium abundance to constrain our parameter space is very mild and for simplicity we opt for not including it. Its small minor impact is explicitly confirmed in the analysis of Ref.~\cite{Froustey:2024mgf}.

Combining 
Eqs.~\eqref{eq:Ypmeasurement} and~\eqref{eq:NeffPlanck} we build a simple Gaussian $\chi^2$. The minimum of $\chi^2$ is $\lesssim 0.1$ and we consider the regions of parameter space to be excluded at 95\% CL for $\Delta \chi^2 > 5.99$ (3.84) when considering two (one) parameter projections. 

\section{Global Analyses: Results}
\label{sec:res}

\begin{figure*}[!t]
\centering
\begin{tabular}{ccc}
\hspace{-0.cm}\includegraphics[width=0.32\textwidth]{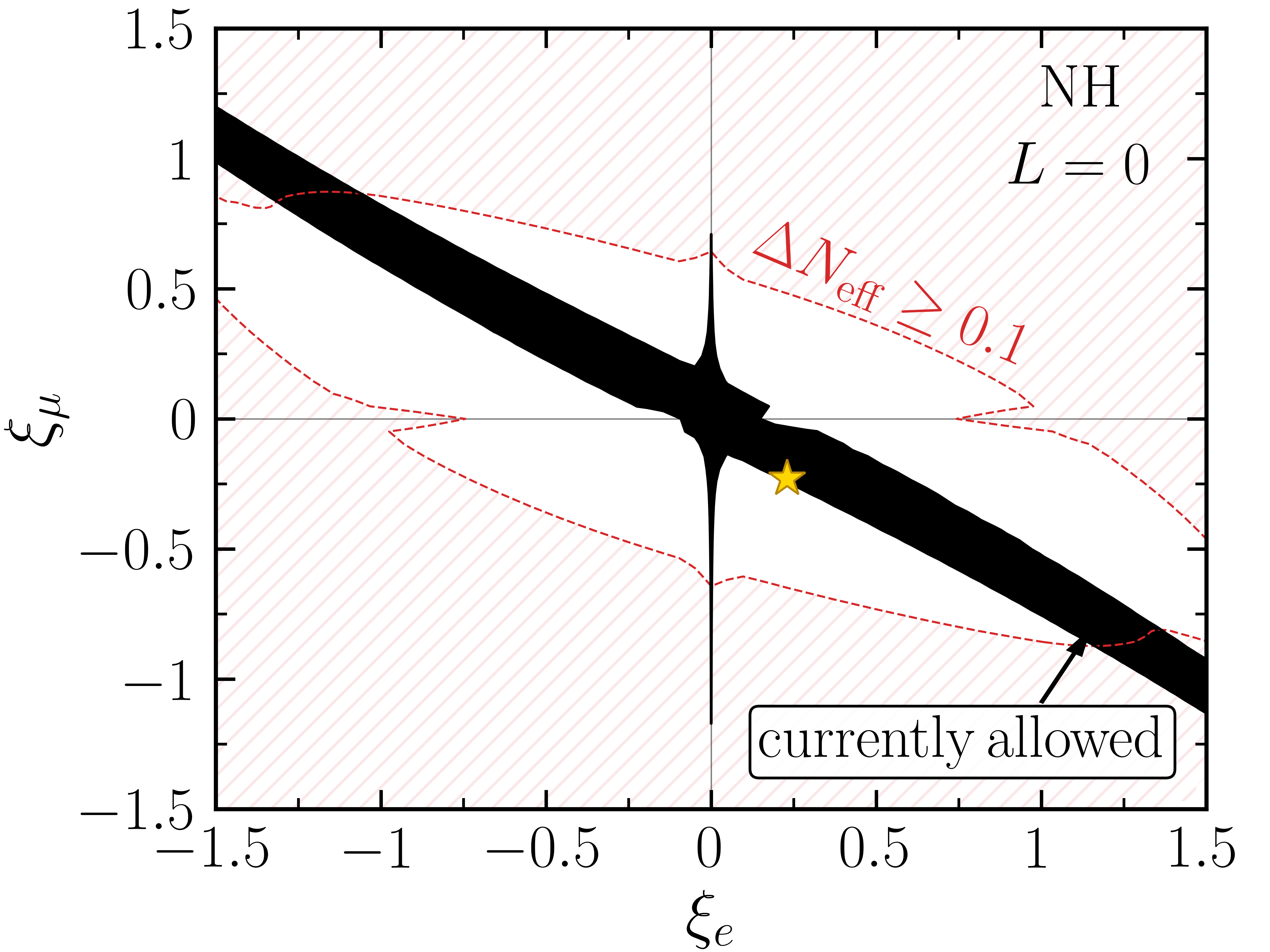}
\hspace{-0.cm}\includegraphics[width=0.32\textwidth]{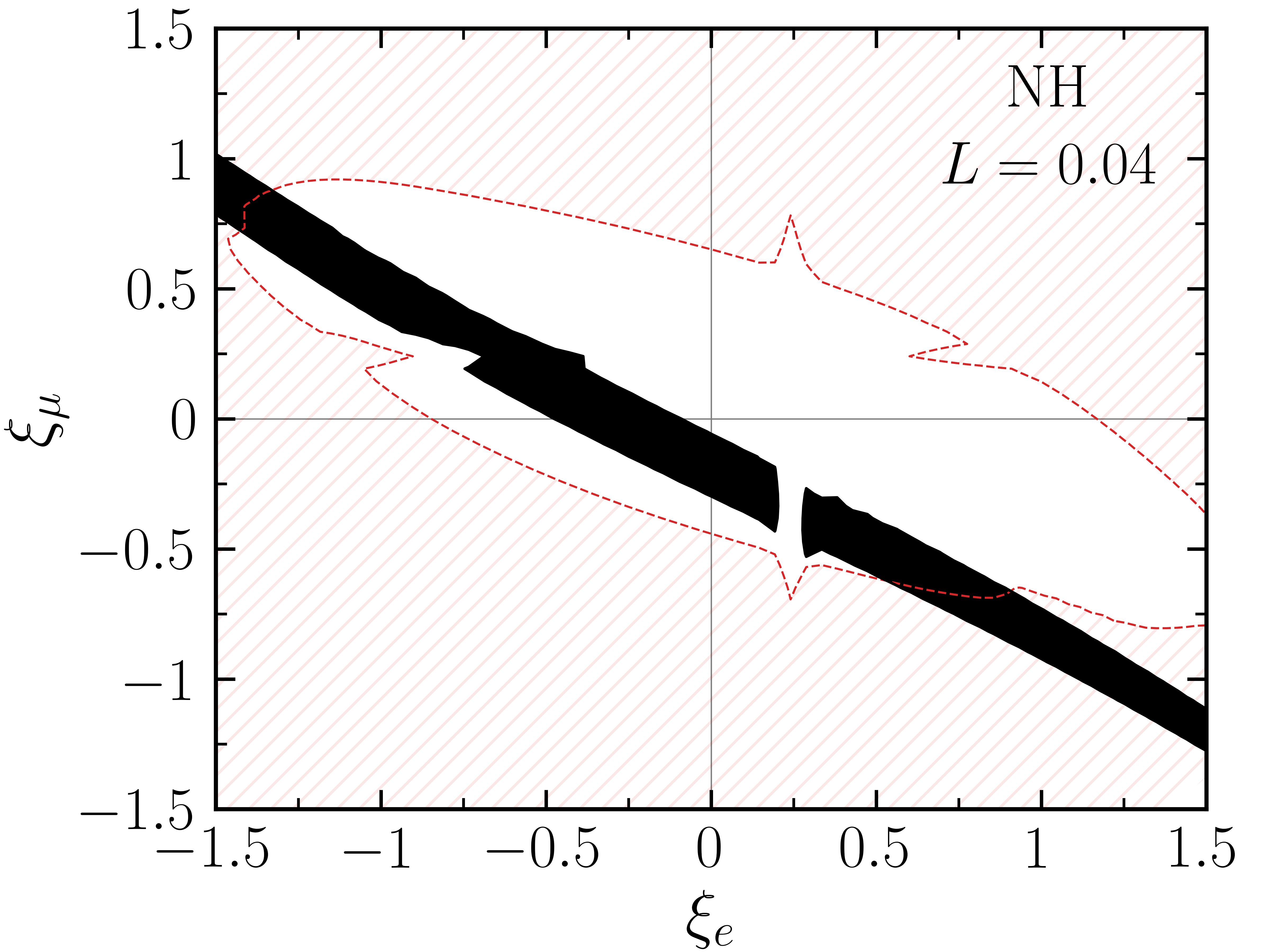}
\hspace{-0.cm}\includegraphics[width=0.32\textwidth]{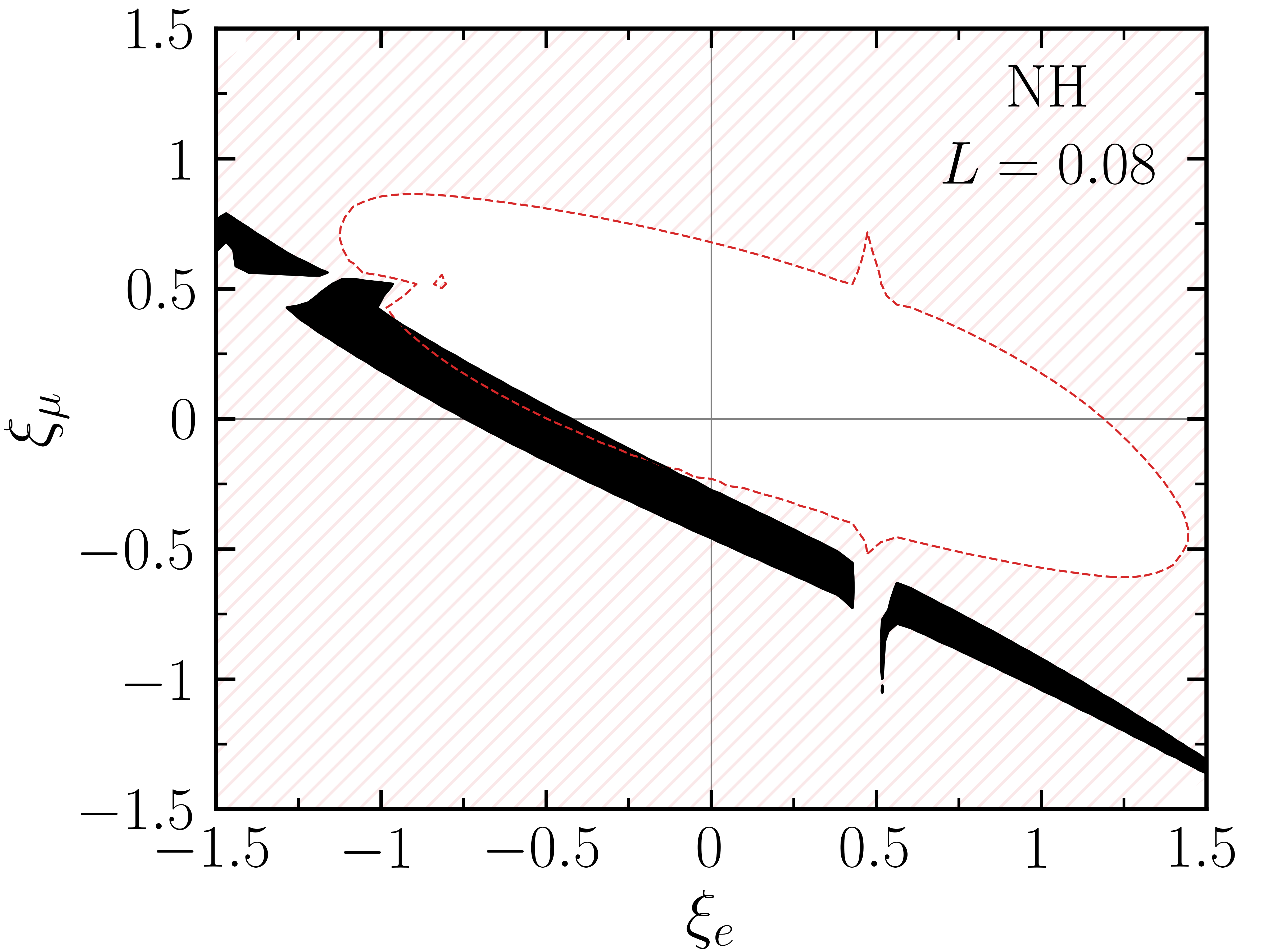} \\
\hspace{-0.cm}\includegraphics[width=0.32\textwidth]{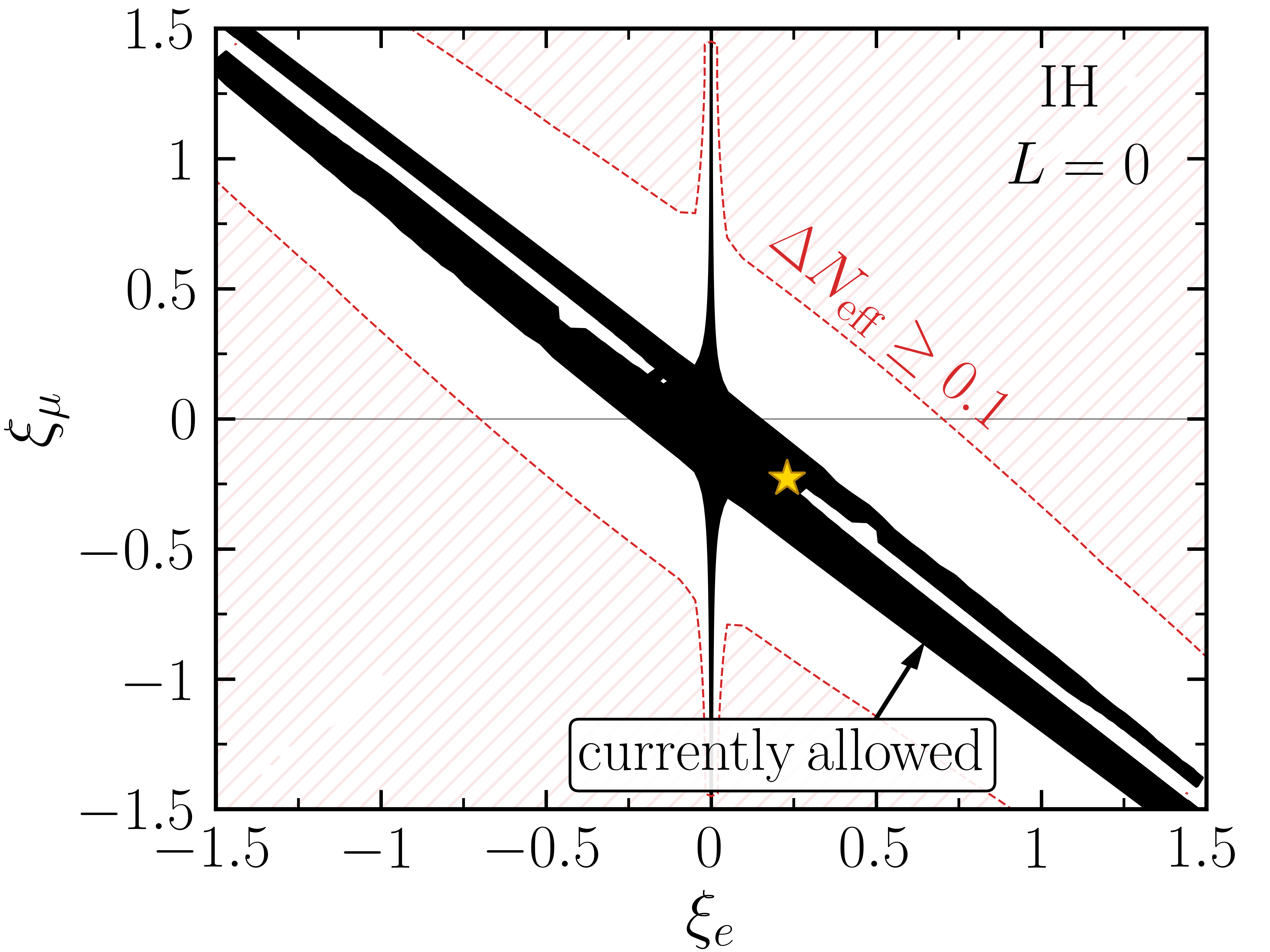}
\hspace{-0.cm}\includegraphics[width=0.32\textwidth]{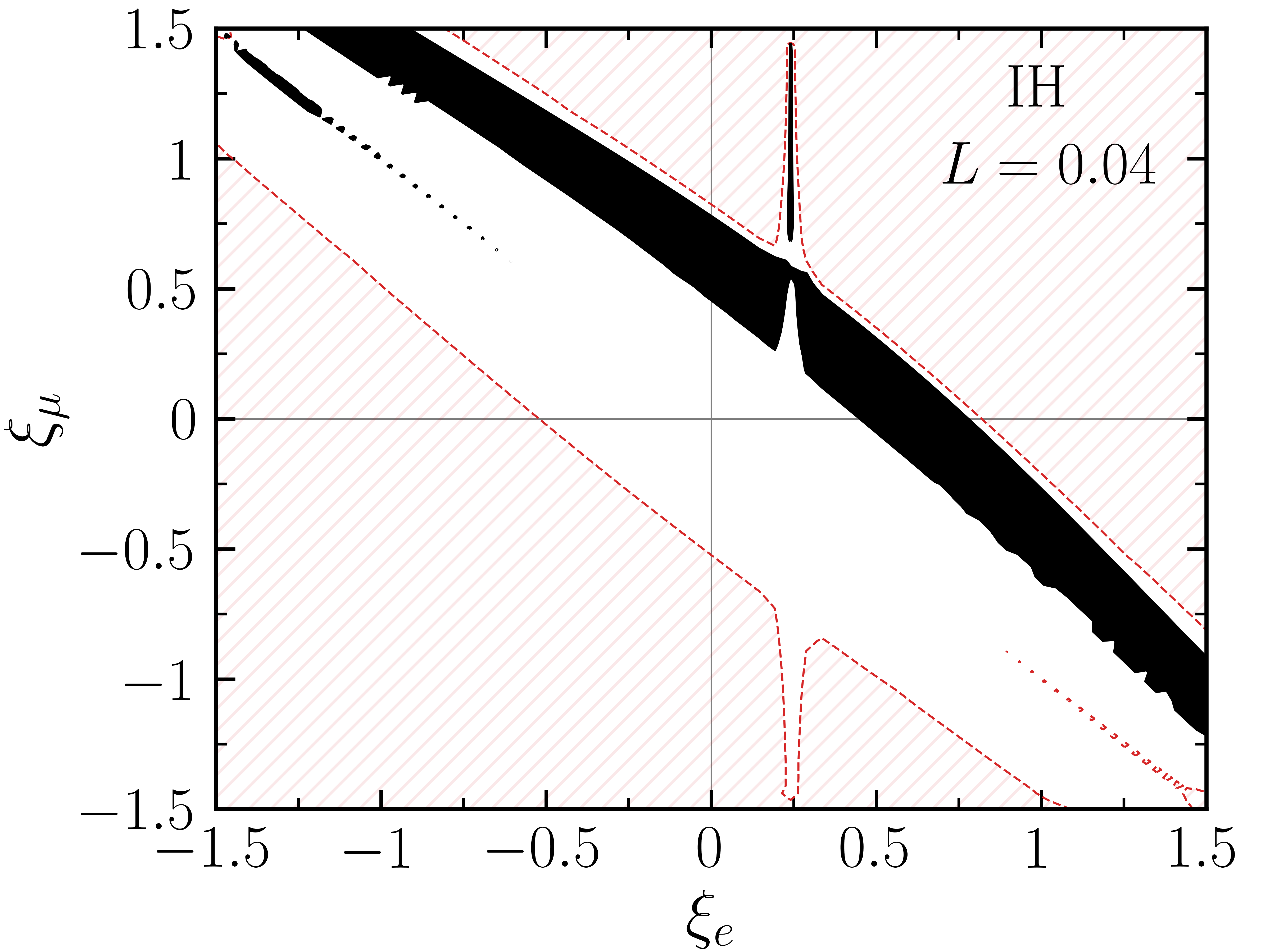}
\hspace{-0.cm}\includegraphics[width=0.32\textwidth]{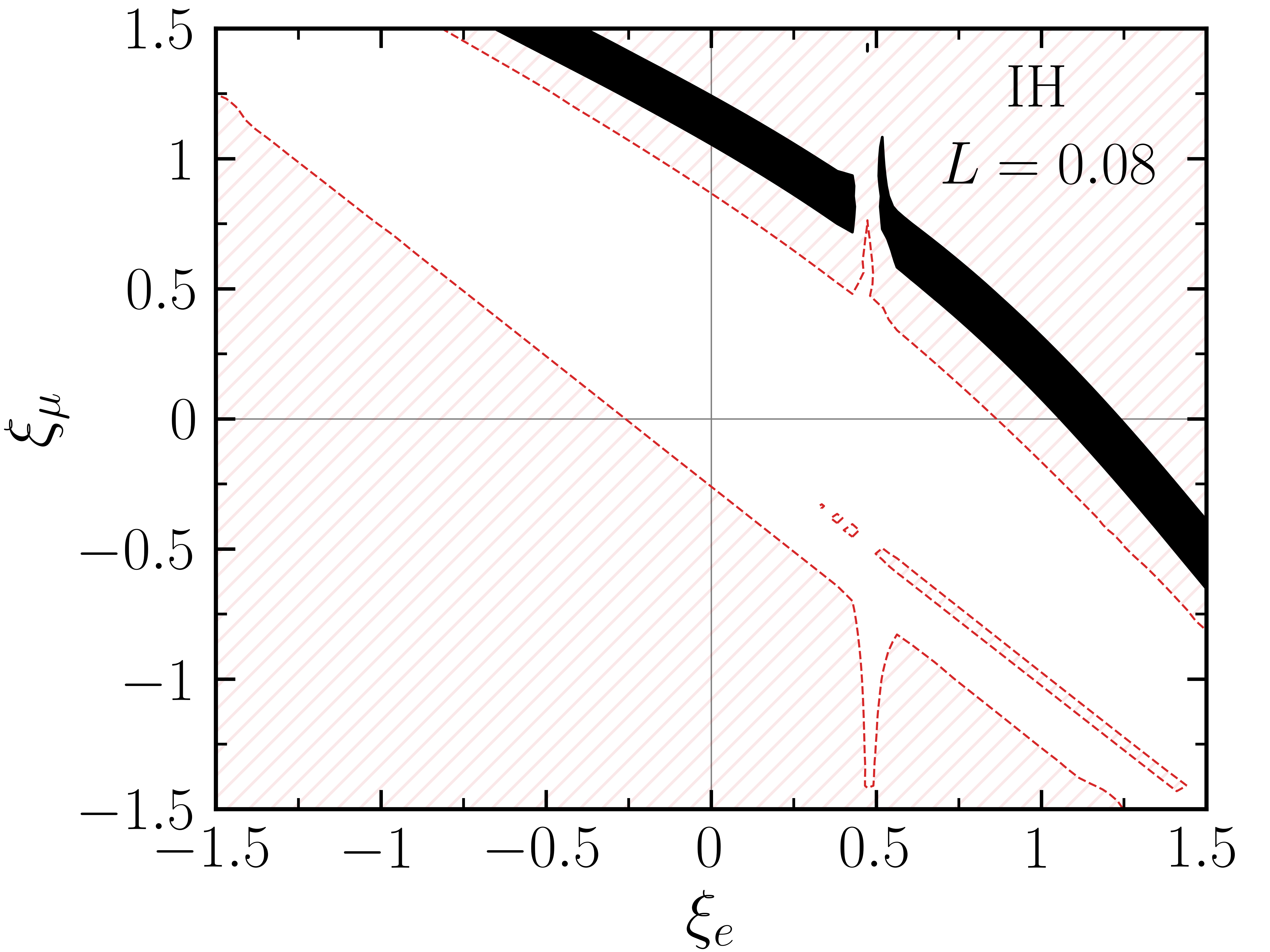} 
\end{tabular}
\caption{
Constraints on primordial lepton flavor asymmetries for increasing total lepton number (from left to right) for normal (upper) and inverted (lower) hierarchy. The black region is the currently allowed region at 95\% CL, $\Delta \chi^2 < 5.99$. The dashed red contours show the improvement expected for the upcoming sensitivity of $\Delta N_{\mathrm{eff}} = 0.1$ centered around the SM value $N_\text{eff} = 3.044$, with the corresponding total lepton number excluded when the black region no longer intersects with the dashed red contour. In the case of $L=0$ we further highlight with a yellow star the scenario of successful generation of the observed baryon asymmetry via tauphobic leptoflavorgenesis, which currently cannot be excluded, see also our previous analysis~\cite{Domcke:2025lzg}.
}
\label{fig:contour-current}
\end{figure*}

\begin{figure*}[!t]
\centering
\begin{tabular}{cc}
\hspace{-0.cm}\includegraphics[width=0.4\textwidth]{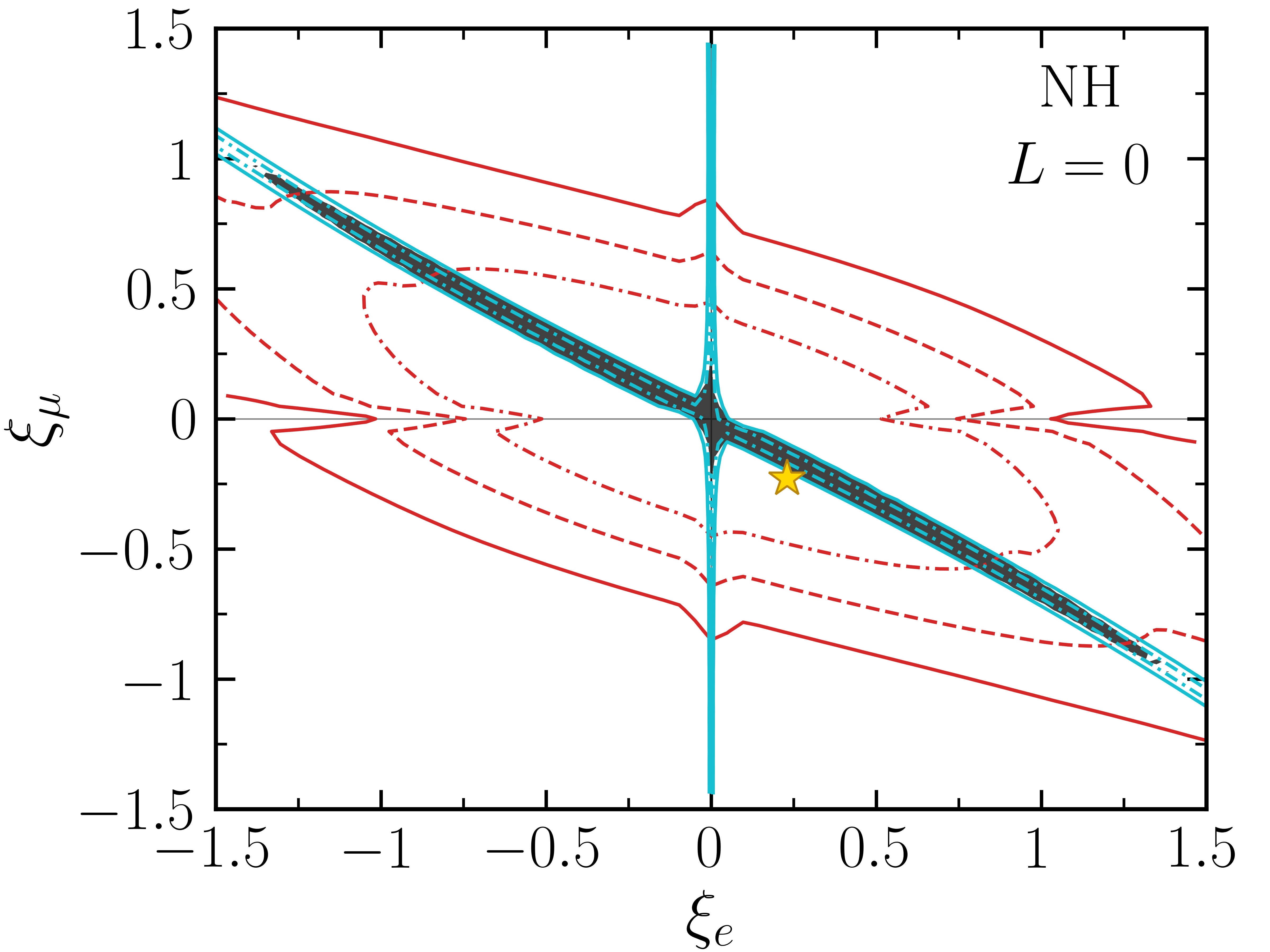}
\hspace{-0.cm}\includegraphics[width=0.4\textwidth]{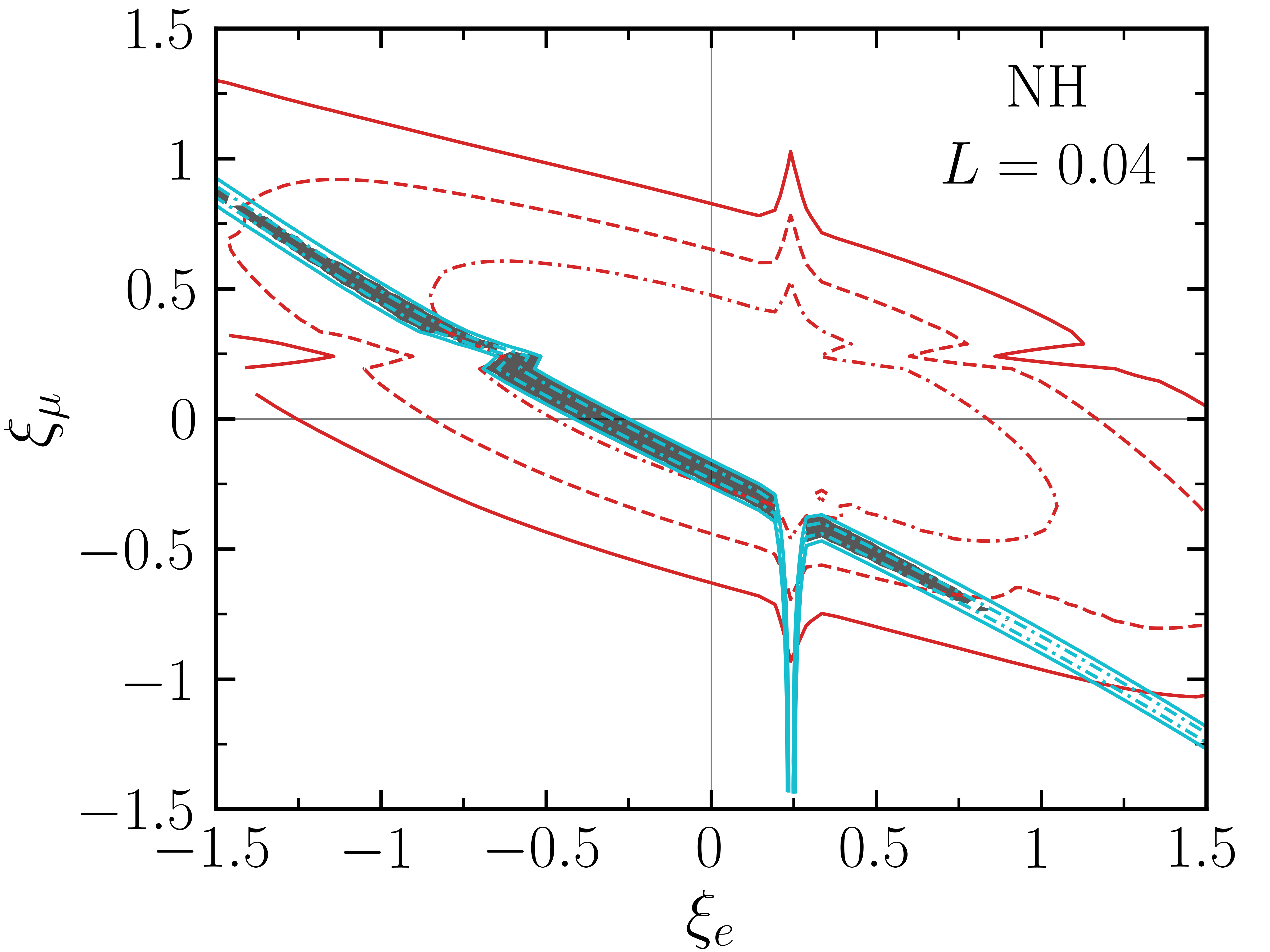}\\
\hspace{-0.cm}\includegraphics[width=0.4\textwidth]{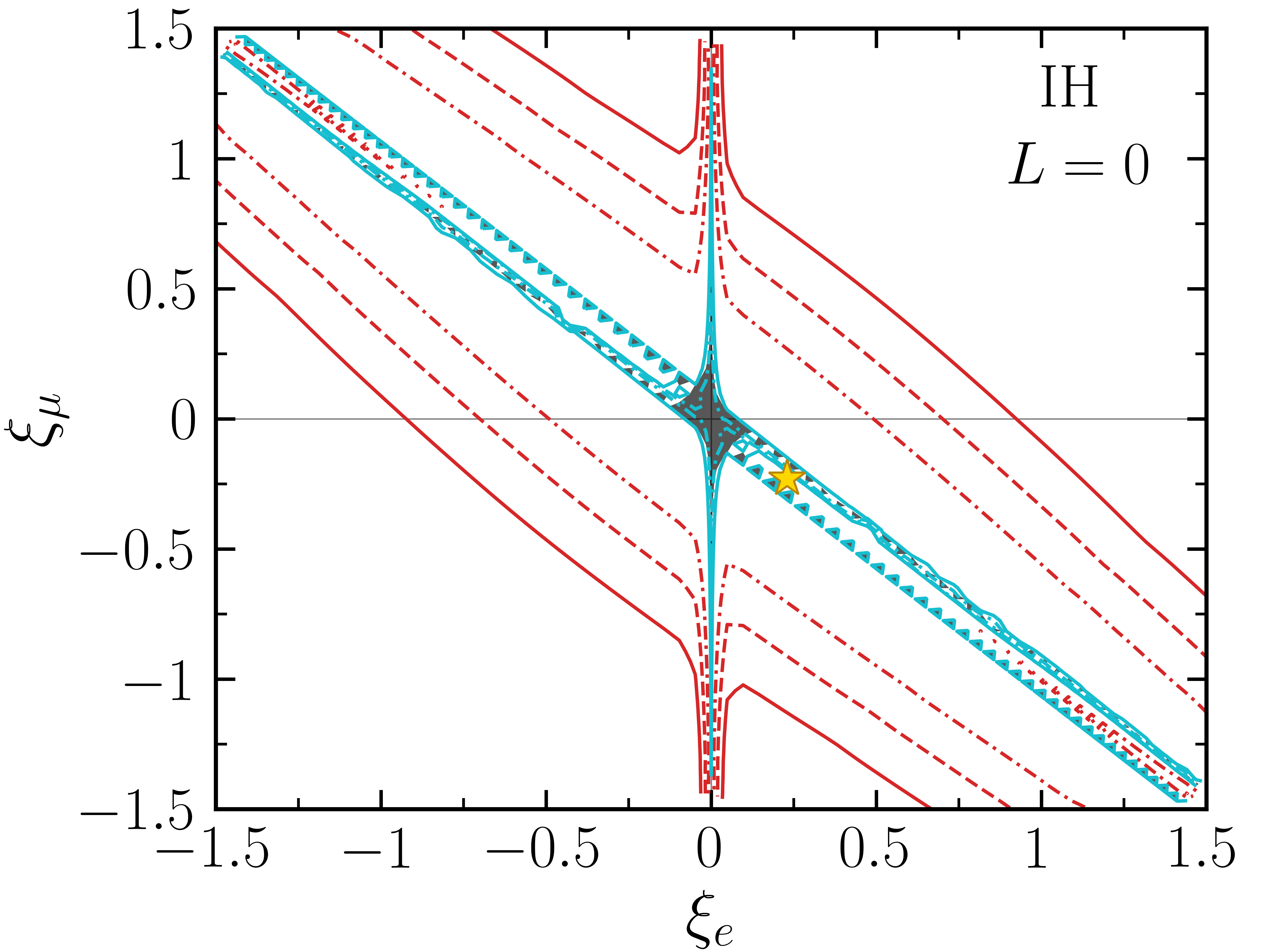}
\hspace{-0.cm}\includegraphics[width=0.4\textwidth]{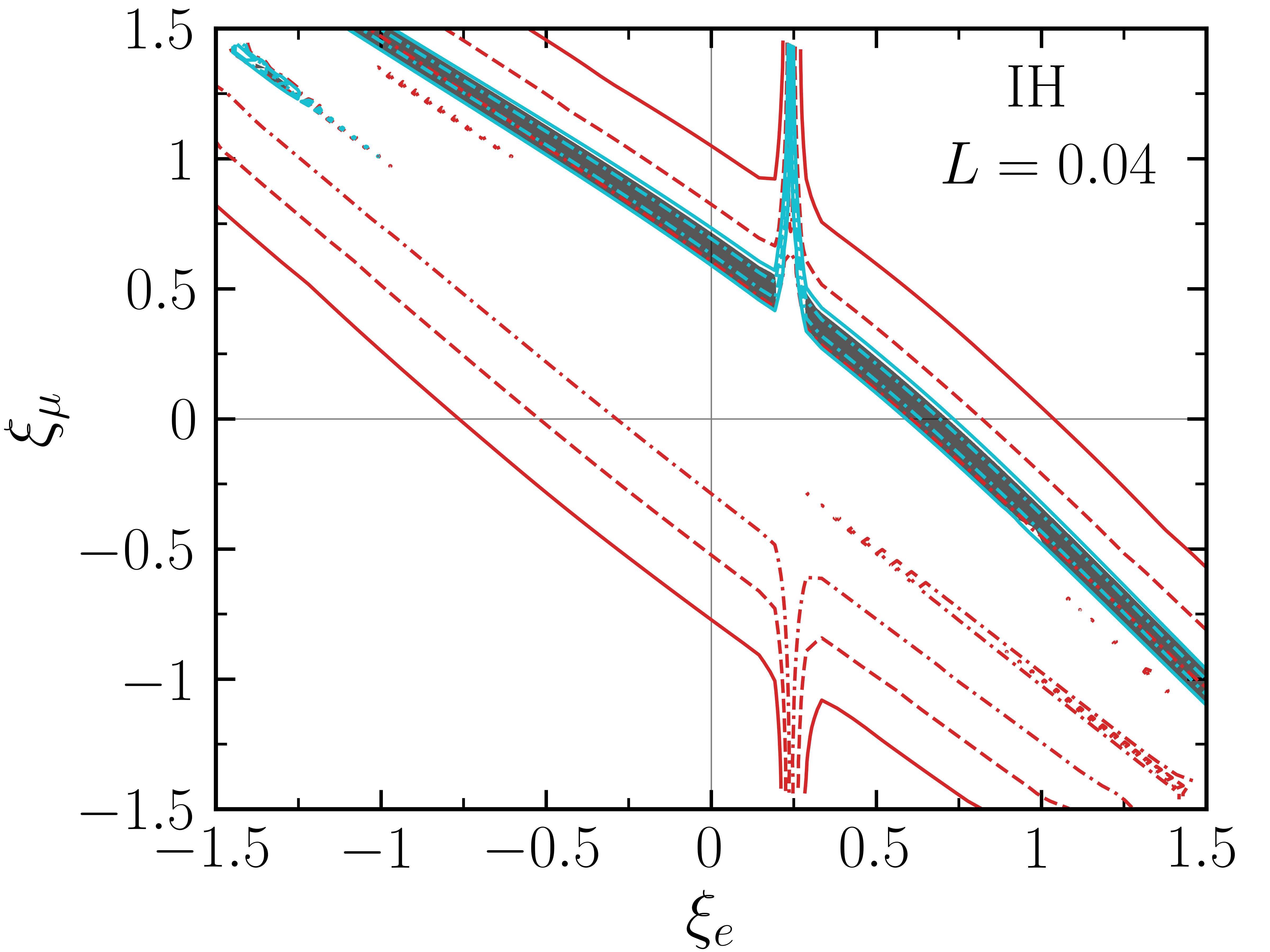}
\end{tabular}
\caption{
Sensitivity to the primordial lepton flavor asymmetries for increasing total lepton number (from left to right) for normal (upper) and inverted (lower) hierarchy.
The light black region is the $\Delta \chi^2 < 5.99$ would be the allowed region assuming a future sensitivity on $Y_P$ of 0.5\% and on $N_{\mathrm{eff}}$ of 0.05 as described in Sec.~\ref{sec:res}. The red lines are contours of $\Delta N_{\mathrm{eff}}$ with solid, dashed and dash-dotted corresponding respectively to $\Delta N_{\mathrm{eff}} = 0.17, 0.1, 0.05$, while the solid (dashed) cyan line represents $\Delta Y_P/Y_{\rm P} = \pm 1.2 \,(\pm 0.5)\%$ centered around the Standard Model value $Y_{\rm P} = 0.247$. See main text for details. 
 }
\label{fig:contour-future}
\end{figure*}
Fig.~\ref{fig:contour-current} summarizes the main results of our study.
It highlights in black for NH and IH the BBN and CMB 95\% CL allowed region as a function of the initial asymmetry in the electron and muon sector. The asymmetry in the tau sector is fixed by choosing a definite total lepton number for each panel, which we set to be $ L = \{0, 0.04, 0.08\}$ from left to right.\footnote{Changing $L \mapsto -L$ while simultaneously changing $\xi_e \mapsto - \xi_e$, $\xi_\mu \mapsto - \xi_\mu$ yields $\xi_\alpha^\text{BBN} \mapsto - \xi_\alpha^\text{BBN}$. The corresponding version of Fig.~\ref{fig:contour-current} essentially is obtained by a reflection on the origin, taking into account that the posterior for $\xi_e^\text{BBN}$ inferred from the observed helium abundance is not perfectly symmetric.} The spikes in vertical and horizontal directions are due to non-adiabatic MSW transitions discussed in App.~\ref{sec:nonadiabatic}, see also~\cite{Domcke:2025lzg}. 

In agreement with our earlier results~\cite{Domcke:2025lzg}, we find that for $L = 0$, the largest lepton flavor asymmetries are allowed for $\xi_e \sim - \xi_\mu$, and at most a very small asymmetry in the tau sector. The present work extends this analysis by i) reducing the theoretical uncertainty due to the identification and exclusion of regions with non-adiabatic MSW transitions (see App.~\ref{sec:nonadiabatic}) and by ii) extending the analysis to non-vanishing total lepton number. As a consequence of the first point, the remaining allowed regions are slightly smaller than in Ref.~\cite{Domcke:2025lzg}, though we note that the particularly interesting case of tauphobic leptogenesis~\cite{March-Russell:1999hpw,Mukaida:2021sgv} remains allowed. The second point, the inclusion of $L \neq 0$, leads to a shift and reduction of the allowed region. The preference for the lower left (upper right) half plane for NH (IH) can be understood from the sign structure of the final value of $\xi_e$ for $L = 0$ (see~\cite{Domcke:2025lzg}), which implies that the addition of positive $L$ shifts the $\xi_e=0$ line in this direction. For $L \neq 0$ the allowed region decreases with increasing total lepton number since total lepton number conservation obstructs the total washout of the lepton flavor asymmetries.

\begin{table}[t]
\begin{center}
\begin{tabular}{|c|c|c|c|}
\hline
\multicolumn{4}{|c|}{\textbf{Bounds/sensitivities on total lepton number}}  \\
\hline  
\hline
\multicolumn{2}{|c|}{Analysis} &  Normal Hierarchy  & Inverted Hierarchy  \\ 
\hline
\multicolumn{2}{|c|}{$2025\,Y_{\rm P}$ + $N_{\rm eff}$} &  $-0.12  \leq  L \leq 0.13$  &$-0.10 \leq  L \leq 0.12$ \\ \hline
\multirow{3}{*}[-12pt]{\;\, \turnbox{90}{Future}\,} 
& $\sigma_{N_{\mathrm{eff}}} = 0.05$  &  $-0.06  \leq  L \leq 0.07 
$  & $-0.05 \leq  L \leq 0.06 $ \\ 
\cline{2-4} & \scalebox{0.75}{$ \frac{\sigma_{Y_{\rm P}}}{Y_{\rm P}}$} $= 0.5$\%     &  $-0.10  \leq L \leq  0.10 $ & $-0.08  \leq L \leq  0.08 $  \\ 
\cline{2-4} & Combined   &  $-0.06 \leq  L \leq 0.07$ & $-0.05 \leq  L \leq 0.06$  \\ \hline
\end{tabular}
\end{center}\vspace{-0.5cm}
\caption{
95\% CL bounds from our analysis on the total lepton number in the Universe $L$ as defined in Eq.~\eqref{eq:L_definition}. The first line corresponds to the current constraint and the other three to sensitivity estimates centered around the corresponding SM values. Improvements on $N_{\rm eff}$ will dominate the sensitivity on $L$, while improved $Y_{\rm P}$ measurements will be particularly useful to constrain the effective electron-neutrino degeneracy parameter, see Fig.~\ref{fig:contour-current} and Fig.~\ref{fig:contour-future} for the bounds as a function of the initial $\nu_e$ and $\nu_\mu$ asymmetries. The improved determination on $N_{\rm eff}$ is expected in $\sim 5$ years while the one for $Y_{\rm P}$ may happen already this year.}
\label{tab:L_bounds}
\end{table}

In Tab.~\ref{tab:L_bounds} we report the current 95\% CL bound on the total lepton number in the Universe:
\begin{subequations}\label{eq:Lbound-current}
\begin{align}
    -0.12< L < 0.13 &\quad   ({\rm NH})\,,\\
    -0.10< L < 0.12 &\quad   ({\rm IH})\,.
\end{align}
\end{subequations}
This is one of the most important results of our study and arises due to a combination of $Y_{\rm P}$ and $N_{\rm eff}$ inferences.  

Tab.~\ref{tab:L_bounds} also shows the impact of improved measurements of both quantities are expected in the near future. 
In particular, we consider a $1\sigma$ sensitivity on $N_{\rm eff}$ of $0.05$ as expected from the Simons Observatory in $\sim 5$ years~\cite{SimonsObservatory:2018koc}. Assuming that the central measured value coincides with the SM value $N_{\rm eff} = 3.044$, this implies an improvement of about a factor of $\sim 2$ in the sensitivity for $L$. In addition, a new high precision measurement of the helium abundance is ongoing using unprecedented high quality spectra for an array of metal poor galaxies~\cite{He4_future}. These measurements are expected to provide a $0.5$\% determination of the helium abundance. Assuming the central value of the measurement will correspond to the Standard Model prediction $Y_{\rm P} = 0.2470$, we find that the limit on $L$ will improve by $\sim 20-30\%$. While this improvement is mild, the sensitivity on the flavor asymmetries will be quite important and we refer to Fig.~\ref{fig:contour-future} for their implications. Finally, combining the two sensitivities one sees from Tab.~\ref{tab:L_bounds} that the sensitivity on $L$ will be dominated by $N_{\rm eff}$ measurements, since this closes the remaining parameter space where the final lepton number is largely stored in $\nu_\mu$ and $\nu_\tau$ neutrinos. This trend is clearly seen in Fig.~\ref{fig:contour-future}.

We remark that the bound on $L$ is rather sensitive to the central value of $N_{\mathrm{eff}}$ and/or $Y_{\rm P}$. In this context, we note that the new results from ACT~\cite{ACT:2025fju,ACT:2025tim} and SPT-3G~\cite{SPT-3G:2025bzu} combined with Planck report a value of $ N_{\rm eff} = 2.81\pm 0.12$ within $\Lambda$CDM. The fact that the mean value is smaller than 3.044 could very well be just a statistical fluctuation and given recent anomalies in several cosmological data sets it is premature to exclude also some systematic effect biasing the central value, but it is clear that a robust measurement of $N_{\rm eff}$ smaller than $3.044$ would strongly disfavor primordial lepton asymmetries as they lead to $\Delta N_{\rm eff} > 0$.

\section{Implications for Today's Universe}

In this section we highlight the consequences of our results for the composition of the Cosmic Neutrino Background (CNB) today. This could have moderate implications for a potential detection of the CNB including the most discussed detection technique of threshold-less neutrino absorption on beta decaying nuclei~\cite{Weinberg:1962zza,Cocco:2007za,PTOLEMY:2018jst,Cheipesh:2021fmg}, while it could impact more significantly other substantially more challenging approaches which have signatures that are directly proportional to the neutrino flavor asymmetries, such as the Stodolsky effect~\cite{Stodolsky:1974aq,Duda:2001hd,Domcke:2017aqj,Bauer:2022lri}.

In the Standard Model the neutrino number density per flavor today is $56/{\rm cm}^{3}$ and the results of our analysis above can be rephrased as 95\% CL bounds on the asymmetries in the number densities of the individual neutrino flavors today,
\begin{subequations}\label{eq:CNBtoday}
\begin{align}
-2 \lesssim    &\, \frac{n_{\nu_e}-n_{{\bar \nu}_e}}{\rm {cm^3}} \lesssim 4 \,,\\
-35 \lesssim    &\, \frac{n_{\nu_\mu}-n_{{\bar \nu}_\mu}}{\rm {cm^3}} \lesssim  35 \,,\\
-30 \lesssim    &\, \frac{n_{\nu_\tau}-n_{{\bar \nu}_\tau}}{\rm {cm^3}} \lesssim  30\,,
\end{align}
\end{subequations}
where the $\nu_e-\bar{\nu}_e$ bound comes mainly from the helium abundance measurement, and where the bounds on the $\mu$ and $\tau$ neutrinos effectively arise from the limit in Eq.~\eqref{eq:Lbound-current} shared by these two neutrinos flavors and hence the bound on the $\mu$ and $\tau$ asymmetries approximately scale as $L/2$. In consequence, these constraints could improve by a factor of $\sim 2$ due to more precise $N_{\rm eff}$ CMB observations, see Tab.~\ref{tab:L_bounds}. We note that while the allowed lepton flavor number density asymmetries in Eq.~\eqref{eq:CNBtoday} can be sizable compared to the Standard Model neutrino number density per flavor, we explicitly checked that the energy density of non-relativistic neutrinos today, $\Omega_{\nu}h^2$, does not change by more than 5\%. Hence, unfortunately, this cannot be used as an additional cosmological constraint. We attribute this lack of sensitivity to the fact that $\Omega_\nu h^2 \propto T_\nu^3$ and $N_{\rm eff} \propto T_\nu^4$, but since $N_{\rm eff}$ is tightly constrained by observations, see Eq.~\eqref{eq:NeffPlanck}, the impact on $\Omega_\nu h^2$ has to be small too.

\section{Summary \& Conclusions}

Based on our framework developed in~\cite{Domcke:2025lzg}, we report constraints on the total lepton number of our Universe as set by BBN and CMB observations. To this end we solve the momentum-averaged neutrino kinetic equations until neutrino decoupling, and develop fitting formulas, see Eq.~\eqref{eq:fittingfuncs}, to match the solution to BBN and CMB observables. A comparison with solutions retaining the full neutrino momentum dependence, demonstrates that our approach is accurate for both the solution of the quantum kinetic equations as well as the prediction of the light element abundances.

\noindent Our main results are:
\begin{enumerate}
    \item The total lepton number in the Universe is bounded by BBN and CMB observations at 95\% CL to
\begin{subequations}
\begin{align}
    -0.12< L < 0.13 &\quad   ({\rm NH})\,, \nonumber \\
    -0.10< L < 0.12 &\quad   ({\rm IH})\,, \nonumber
\end{align}
\end{subequations}
where $L $ is defined in Eq.~\eqref{eq:L_definition} and which approximates to $L\simeq (n_\nu-n_{\bar{\nu}})/T_\nu^3$ and is related to the yield by $Y_L \simeq L/4.7$. 
\item The primordial flavor lepton asymmetries currently compatible with BBN and CMB observations are depicted in Fig.~\ref{fig:contour-current}. In particular, from the left panels we note that the parameter points where baryogenesis can be explained by large and compensated lepton asymmetries in the $e$ and $\mu$ flavors~\cite{March-Russell:1999hpw,Mukaida:2021sgv} are  still allowed although close to being excluded. 
    \item We have explored the sensitivity of ongoing CMB experiments and campaigns to measure $Y_{\rm P}$ with increased precision. We find that upcoming $N_{\rm eff}$ measurements have the potential to improve the sensitivity on $L$ by a factor of $\sim 2$ and that improved $Y_{\rm P}$ determinations can also contribute to improvements on $L$ but will be particularly useful in constraining regions of the primordial lepton asymmetries, see Fig.~\ref{fig:contour-future}. 
\item Our results bound the composition of the Cosmic Neutrino Background today according to Eq.~\eqref{eq:CNBtoday}.
\end{enumerate}
Our limits are complementary and independent to those arising from baryon number generation at the electroweak phase transition and apply also  to lepton number generation mechanisms at $T < 130\,{\rm GeV}$. Our new bounds have implications for the maximal production of sterile neutrino dark matter~\cite{Shi:1998km,Shaposhnikov:2023hrx,Akita:2025txo,Akita:2025txo,Vogel:2025aut,Kasai:2025xaw}, and the possibility of a QCD 1st order phase transition~\cite{Gao:2021nwz,Gao:2023djs,Gao:2024fhm}. 

We highlight that we publicly release the C++ code \texttt{COFLASY-C} on~\gitlink $\,$ which solves the neutrino lepton flavor asymmetry evolution on a $\mathcal{O}({10})\,{\rm s}$ timescale. In this repository we also provide all the relevant results from our study (including the data used to generate Fig.~\ref{fig:contour-current} and Fig.~\ref{fig:contour-future}). 

Future measurements of the primordial helium abundance~\cite{He4_future}, refinements in the predictions of the deuterium abundance~\cite{Pitrou:2021vqr}, and upcoming measurements of $N_{\rm eff}$ from CMB observations~\cite{SimonsObservatory:2018koc} will be able to better constrain the parameter space for primordial lepton asymmetries. We have demonstrated that our framework will be accurate to interpret these future measurements.

\vspace{0.5 cm}
\begin{center}
\textbf{Acknowledgments}
\end{center}

We thank Julien Froustey and Cyril Pitrou for helpful discussions on solving the quantum kinetic equations. We are particularly grateful to Julien Froustey for generously providing comparison plots using their momentum-dependent code \texttt{NEVO}, which have proven extremely useful in understanding various aspects of the system.
\noindent
M.F.N. and S.S. would like to thank the CERN Theory group for hospitality and financial support  while part of this work was performed. We acknowledge support from the DOE Topical Collaboration ``Nuclear Theory for New Physics,'' award No.~DE-SC0023663. M.F.N. is supported by the STFC under grant ST/X000605/1. S.S. was supported by the US Department of Energy Office and by the Laboratory Directed Research and Development (LDRD) program of Los Alamos National Laboratory under project numbers 20230047DR and 20250164ER. Los Alamos National Laboratory is operated by Triad National Security, LLC, for the National Nuclear Security Administration of the U.S. Department of Energy (Contract No. 89233218CNA000001). We gratefully acknowledge the computer resources of the \textit{lxplus} cluster, funded and operated by CERN.

\appendix
\addtocontents{toc}{\vspace{+0.5em}}  
\section{Non-Adiabatic MSW Transitions}\label{sec:nonadiabatic}

\begin{figure*}[!t]
\centering
\begin{tabular}{cc}
\hspace{-0.cm}\includegraphics[width=0.45\textwidth]{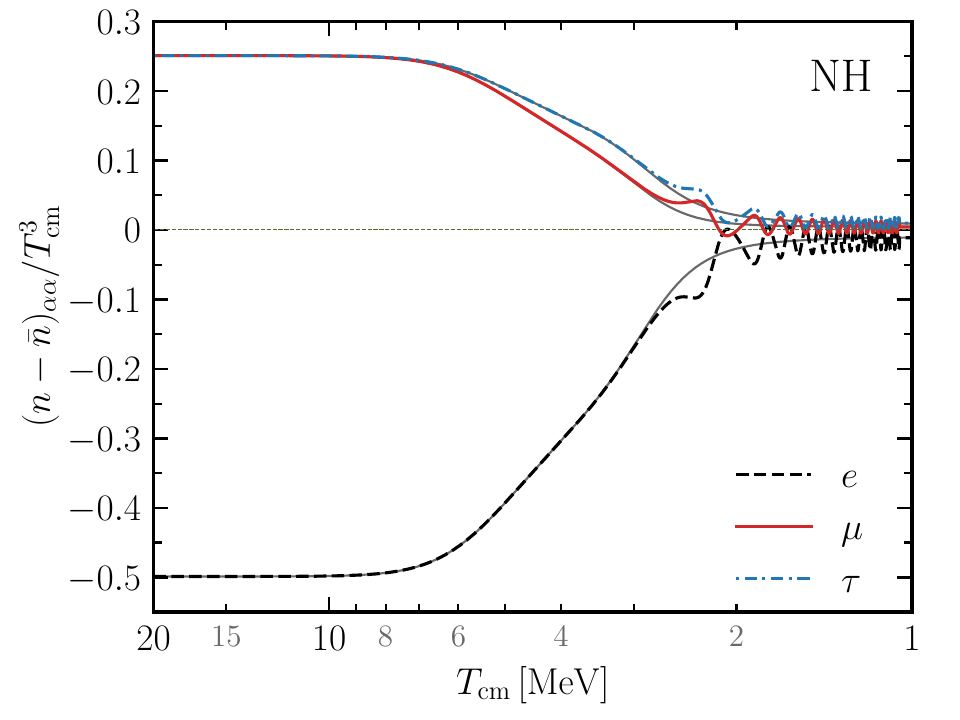}
\hspace{-0.cm}\includegraphics[width=0.45\textwidth]{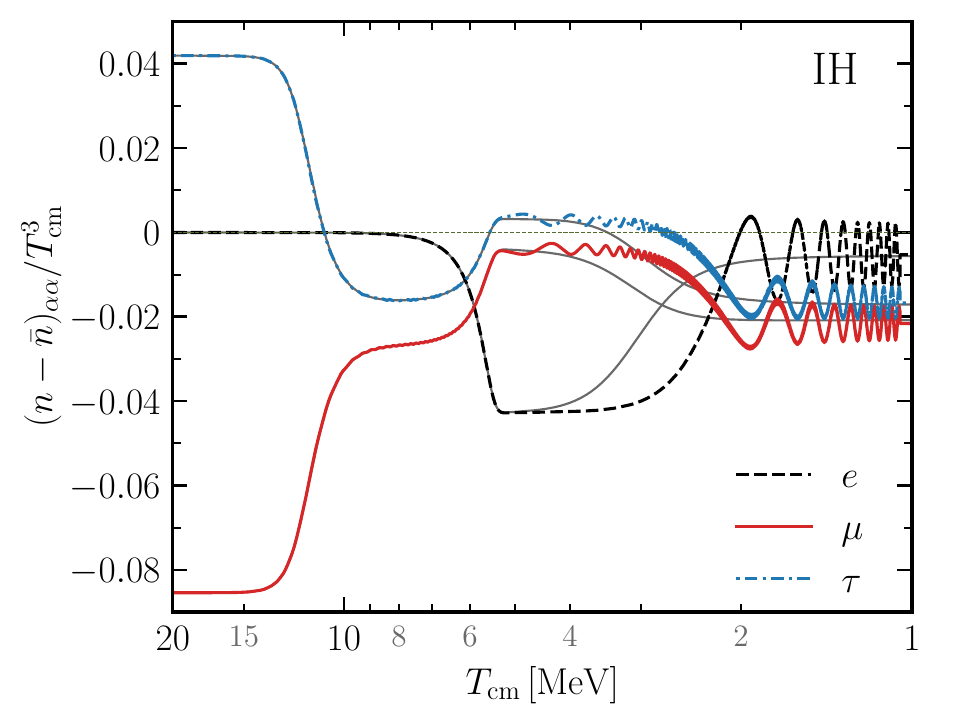}\\
\hspace{-0.cm}\includegraphics[width=0.45\textwidth]{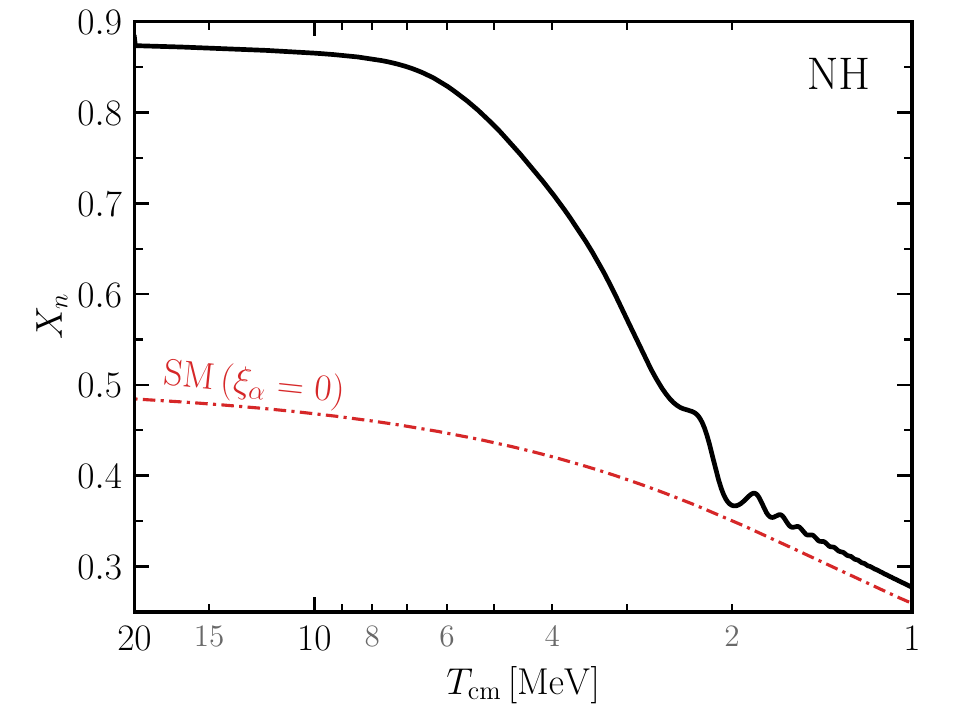}
\hspace{-0.cm}\includegraphics[width=0.45\textwidth]{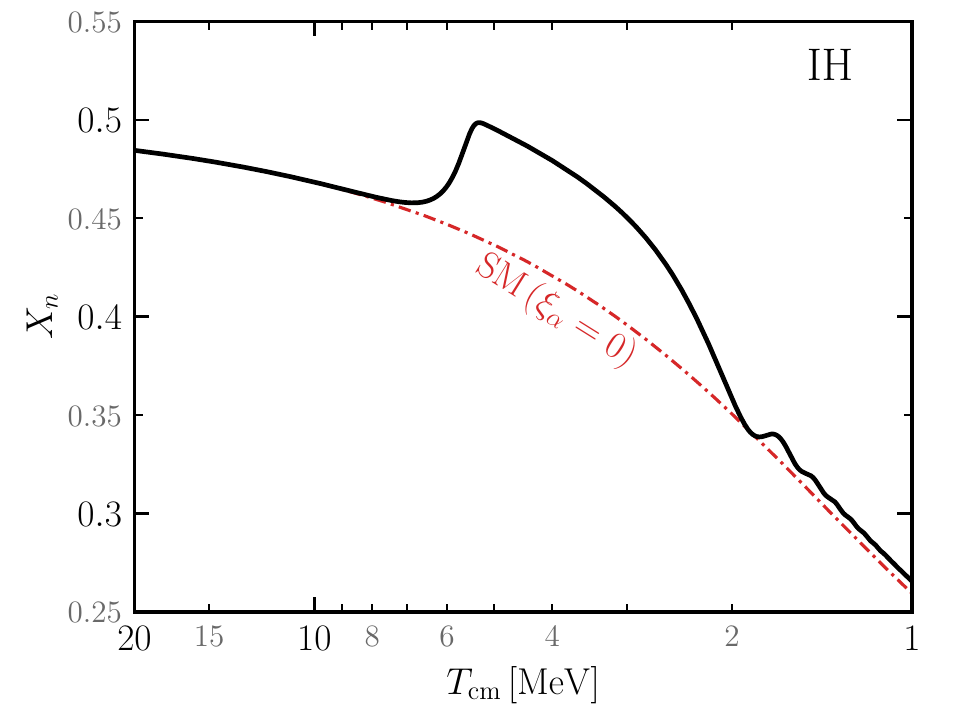}
\end{tabular}
\caption{
Time evolution plots for the lepton asymmetry (first row) and the free neutron fraction (second row) for NH (left) and IH (right) in the presence of a non-adiabatic second electron-driven MSW transition.
Grey lines in the lepton asymmetry evolution represent the adiabatic solution ($V_s = 0$). 
We observe a significant deviation in the time evolution, but asymptotically it (almost) coincides with the full non-adiabatic solution.
Even in the case of late time oscillations with significant amplitude, we see that they are transferred only very mildly into the $X_n$ evolution, further justifying our approach to assume an adiabatic $\xi_e$ evolution at $T < 1.5\,\mathrm{MeV}$, see also Sec.~\ref{subsec:BBN_Analysis}.
 }
\label{fig:2eMSW}
\end{figure*}

At high temperatures, the vacuum Hamiltonian ${\cal H}_0$ is negligible, and the evolution of the neutrino density matrices is governed by the matter potentials ${\cal V} = V_c + V_s$ and the collision terms ${\cal I}$, which are diagonal in flavor space. Mikheyev-Smirnov-Wolfenstein (MSW) transitions occur when the off-diagonal terms in the vacuum Hamiltonian, controlled by the mass splittings, come to dominate over the charged current contributions contained in $V_c$. They cause the onset of neutrino oscillations. MSW transitions can be classified by the flavor of the element in $V_c$ and the mass splitting in ${\cal H}_0$ which match at the transitions: we distinguish electron- and muon driven MSW transitions, as well as first ($\Delta m_{31}^2$ - driven) and second ($\Delta m_{21}^2$-driven) transitions.

As discussed in Refs.~\cite{Froustey:2021azz,Domcke:2025lzg}, in most of the parameter space, the MSW transitions are adiabatic, i.e.\ the change in the Hamiltonian ${\cal H}$ is slow compared to the frequency of synchronous oscillations. The latter describe the aligned precession of the asymmetry vectors of all momentum modes around the Hamiltonian in $SU(N)$ space~\cite{Pastor:2001iu,Dolgov:2002ab,Abazajian:2002qx,Wong:2002fa,Mangano:2010ei,Froustey:2021azz}, and are due to the domination of the $V_{s}$ term in the QKEs throughout the evolution. In this case, the density matrices become aligned with $V_{s}$ (i.e.\ the commutators $[V_{s},\rho] \simeq [V_{s},\bar{\rho}]\simeq 0$ vanish) and thus to good approximation one can set $V_s=0$ in the QKEs. This is what we refer to as adiabatic approximation, see also Ref.~\cite{Domcke:2025lzg}. However, in particular flavor directions, the frequency of synchronous oscillations can become anomalously small, resulting in non-adiabatic MSW transitions and consequently a non-adiabatic evolution of the number densities $n_\alpha$. In Ref.~\cite{Domcke:2025lzg} this was investigated for the (first) muon-driven MSW transition and the first electron-driven MSW transition, finding good agreement between analytical estimations of the synchronous oscillation frequency, the momentum-averaged numerical results, and the full momentum-dependent numerical results of~\cite{Froustey:2021azz,Froustey:2024mgf}. The presence of non-adiabatic MSW transitions significantly impacts the final neutrino asymmetries, leading to specific flavor directions in which the asymmetry washout is particularly strong or weak. In particular, the non-adiabatic muon-driven MSW occurring for $ \xi_\mu = - \xi_\tau$ (i.e.\ $\xi_e = \xi_\text{tot}$) decreases the final asymmetry while the first non-adiabatic electron MSW occurring for $ \xi_e = -  \xi_\tau$ (i.e.\ $ \xi_\mu =  \xi_\text{tot}$) in NH and for $ \xi_e = -  \xi_\mu$ (i.e.\ $ \xi_\mu = -  \xi_e$) in IH leads to larger remaining asymmetry~\cite{Domcke:2025lzg}. 

In this appendix, we point out the role of a non-adiabatic second electron-driven MSW transition, which has not been investigated so far. The transition occurs when the off-diagonal elements of the vacuum Hamiltonian proportional to $\Delta m_{21}^2$ come to dominate over $(V_c)_{11}$, which occurs around $3\,\mathrm{MeV}$. At this point in time, the asymmetries have already dropped significantly compared to their initial values. We observe non-adiabatic second electron-driven MSW transitions in NH for e.g.\ $\xi_\mu \simeq \xi_\tau$ when $L \ll |\xi_\alpha|$, see left panels of Fig.~\ref{fig:2eMSW}. Contrary to the non-adiabatic MSW transitions discussed in Refs.~\cite{Froustey:2021azz,Domcke:2025lzg}, in this case we can not assign this phenomenon to an anomalously slow frequency of the synchronous oscillations, but instead we associate it with a particularly rapid change in the Hamiltonian, due to the rapid washout of asymmetries.

A comparison with the momentum-dependent code \texttt{NEVO}~\cite{Froustey:2020mcq,Froustey:2021azz}\footnote{Special thanks to Julien Froustey for generously providing comparison plots obtained with \texttt{NEVO}.} reveals that this phenomenon is qualitatively present and very similar both with and without the momentum average. However, since the non-adiabatic evolution entails a rather complex evolution pattern of the asymmetries, the final asymmetries are particularly sensitive to small errors introduced by the momentum averaging. This explains the larger discrepancy between the two approaches in specific flavor directions, noted in~\cite{Domcke:2025lzg}. The results presented in this paper, focusing on sizable total lepton number, are not subject to this uncertainty, see also App.~\ref{sec:BBNvalidation}.

Finally, we point to a second class of non-adiabatic second electron-driven MSW transitions which we observe for $\xi_\tau = -\sum \xi_\alpha$ in the IH case. Here, the evolution becomes anomalously slow and changes in the matter potential appear rapid in comparison, similar to the non-adiabatic MSW transitions discussed in Refs.~\cite{Froustey:2021azz,Domcke:2025lzg}. We depict such an example in the right panels of Fig.~\ref{fig:2eMSW}. In this case, a comparison with \texttt{NEVO} shows good agreement.

\vspace{-0.3cm}

\section{Proton-Neutron Interaction Rates}\label{sec:np_pn_rates}

Here, for completeness, we summarize the neutron-proton conversion rates assuming heavy baryon chiral perturbation theory. This effectively amounts to assuming that neither neutron nor proton experience any recoil, which sometimes is referred to as the infinite nucleon mass approximation. In this limit, these rates can be written as, see e.g.~\cite{Sarkar:1995dd}:
\begin{subequations}
    \begin{align}
    \label{eq:BornRates_forward}
    \begin{split}
    \lambda_{n+\nu_e\to p +e^-} &= A \int_0^{\infty} \dd k_\nu k_\nu^2 k_e E_e  f_\nu(1-f_e)   \\ 
    &\qquad  \text{with} \,\,\,E_e = E_\nu +\Delta m \,, 
    \end{split}
    \\
    \begin{split}
    \lambda_{n+e^+\to p + \bar{\nu}_e} &= A \int_0^{\infty} \dd k_e k_\nu^2 k_e^2  f_{\bar{e}}(1-f_{\bar \nu_e})  \\ 
    &\qquad \text{with} \,\,\,E_\nu = E_e +\Delta m \,,
    \end{split}
    \\
    \begin{split}
    \lambda_{n\to p + e^- + \bar{\nu}_e} &= A  \int_0^{\sqrt{\Delta m^2-m_e^2}} \!\!\!\!\!\!\!\!\!\!\!\!\!\!\!\!\!\!  \dd k_e k_\nu^2 k_e^2 (1- f_{\bar{\nu}_e})(1-f_{e})   \\
    &\qquad \text{with} \,\,\,E_\nu = \Delta m-E_e  \,,
    \end{split}
\end{align}
\end{subequations}
where $\Delta m = 1.293\,\mathrm{MeV}$ is the proton neutron mass splitting and, since neutrinos can be treated as massless, we take $k_\nu = E_\nu$. The constant $A$ is obtained by matching the last rate to the neutron decay rate in the vacuum limit, i.e. $f_{\bar{\nu}_e},f_e \mapsto 0$. One finds:
\begin{align}
    A &= \frac{1}{\tau_n} \frac{1}{0.0157527 \Delta m^5} \,,
\end{align}
where $\tau_n$ is the neutron lifetime. The PDG reports an average of $\tau_n = 878.4\pm 0.5\,{\rm s}$~\cite{ParticleDataGroup:2024cfk} and since the error has a tiny impact on the primordial abundances we fix $\tau_n = 878.4\,{\rm s}$ in our calculations. 

We also need the inverse reaction rates which read:
\begin{subequations}
    \begin{align}
    \label{eq:BornRates_backward}
    \begin{split}
    \lambda_{p + e^-\to n+\nu_e} &= A \int_{\sqrt{\Delta m^2-m_e^2}}^{\infty} \dd k_e k_\nu^2 k_e^2  f_e(1-f_{\nu_e})  \\ 
    &\qquad \text{with} \,\,\,E_\nu = E_e -\Delta m \,,
    \end{split}
    \\
    \begin{split}
    \lambda_{p + \bar{\nu}_e\to n+e^+} &=   A \int_{\Delta m+m_e}^{\infty} \dd k_\nu k_\nu^2 k_e E_e  f_{\bar{\nu}_e}(1-f_{\bar e})  \\ 
    &\qquad \text{with} \,\,\,E_e = E_\nu -\Delta m \,, 
    \end{split}
    \\
    \begin{split}
    \lambda_{p + e^- + \bar{\nu}_e \to n} &= A  \int_0^{\sqrt{\Delta m^2-m_e^2}}  \dd k_e k_\nu^2 k_e^2 f_e f_{\bar{\nu}_e} \\
    &\qquad \text{with} \,\,\,E_\nu = \Delta m-E_e \,.
    \end{split}
\end{align}
\end{subequations}
Radiative corrections and the effect of nucleon recoil amount to a $\lesssim 2\%$ effect on top of these rates, which as explained in the main text, we take as a universal multiplicative factor as a function of $T_\gamma$ as calculated in the Standard Model using \texttt{PRIMAT}~\cite{Pitrou:2018cgg}.

\section{Validation of the BBN framework }\label{sec:BBNvalidation}

In this appendix we show explicit comparison between the results from our solutions for the primordial element abundance of helium as compared with the full solutions from Ref.~\cite{Froustey:2024mgf} which solved the momentum-dependent quantum kinetic equations and calculated the primordial element abundances using \texttt{PRIMAT} for selected parts of the parameter region.

\begin{figure*}[!t]
\begin{tabular}{ccc}
\hspace{-1.66cm} \includegraphics[width=0.44\textwidth]{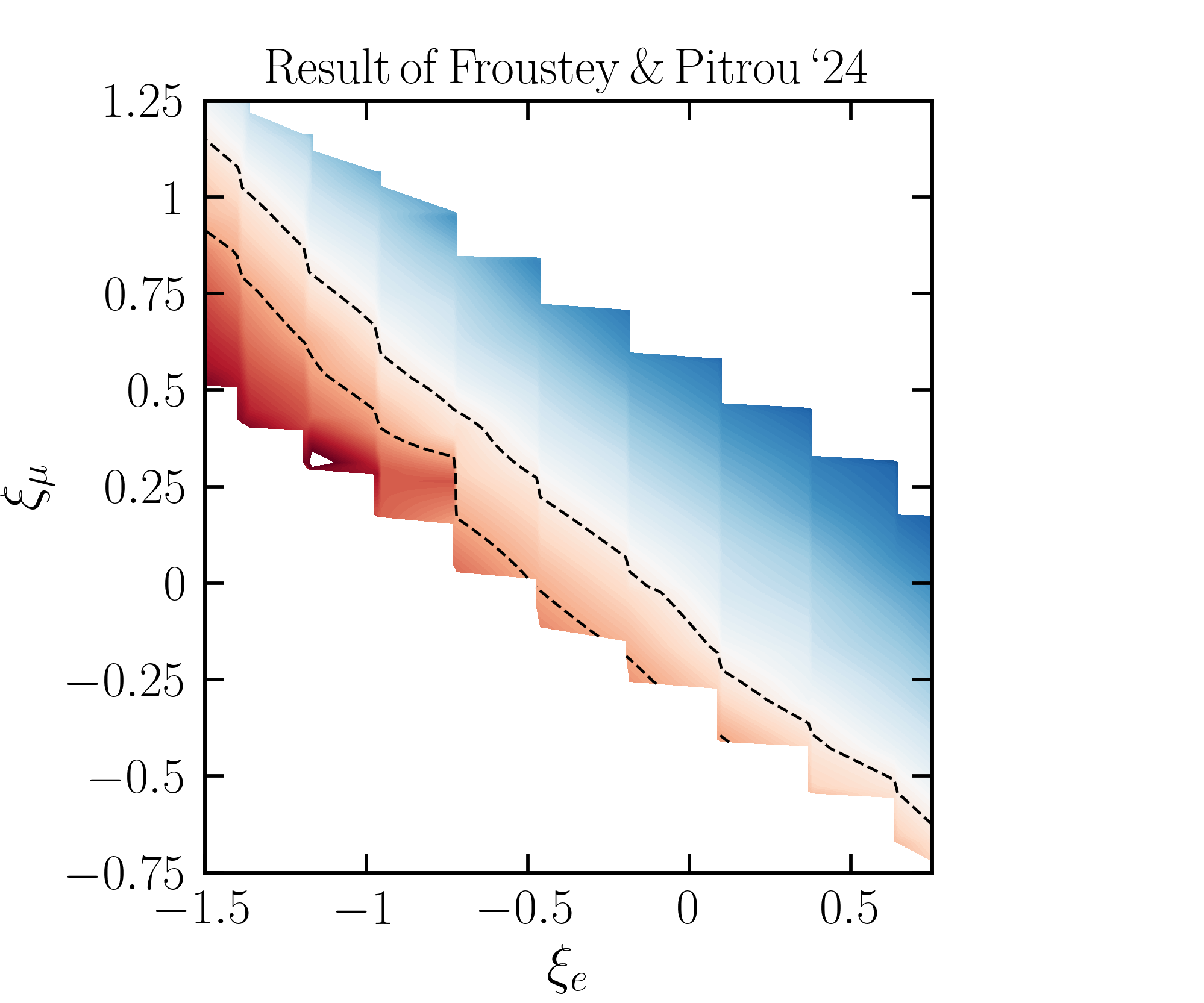} 
& \hspace{-1.75cm}  \includegraphics[width=0.44\textwidth]{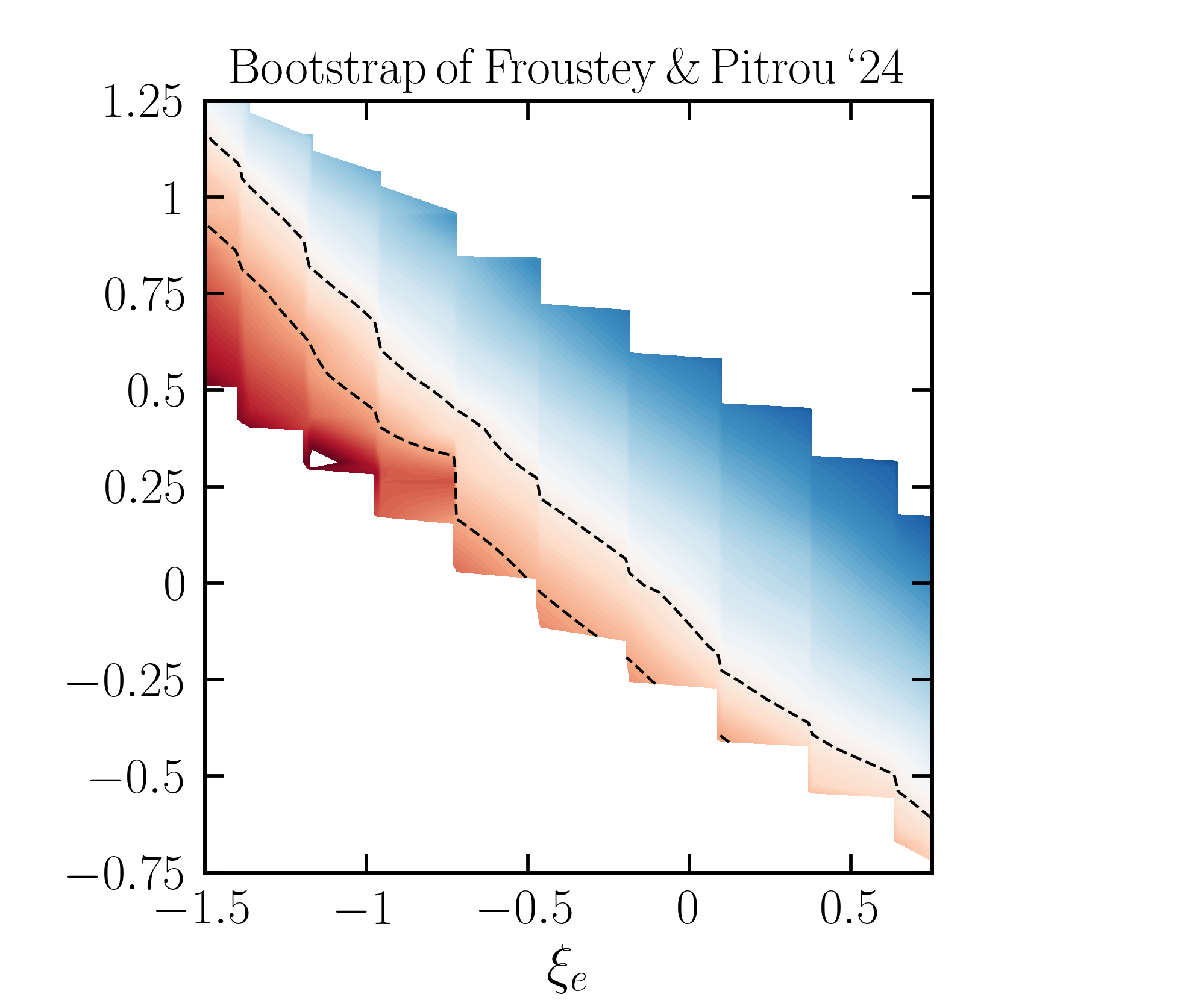}
& \hspace{-1.75cm}  \includegraphics[width=0.44\textwidth]{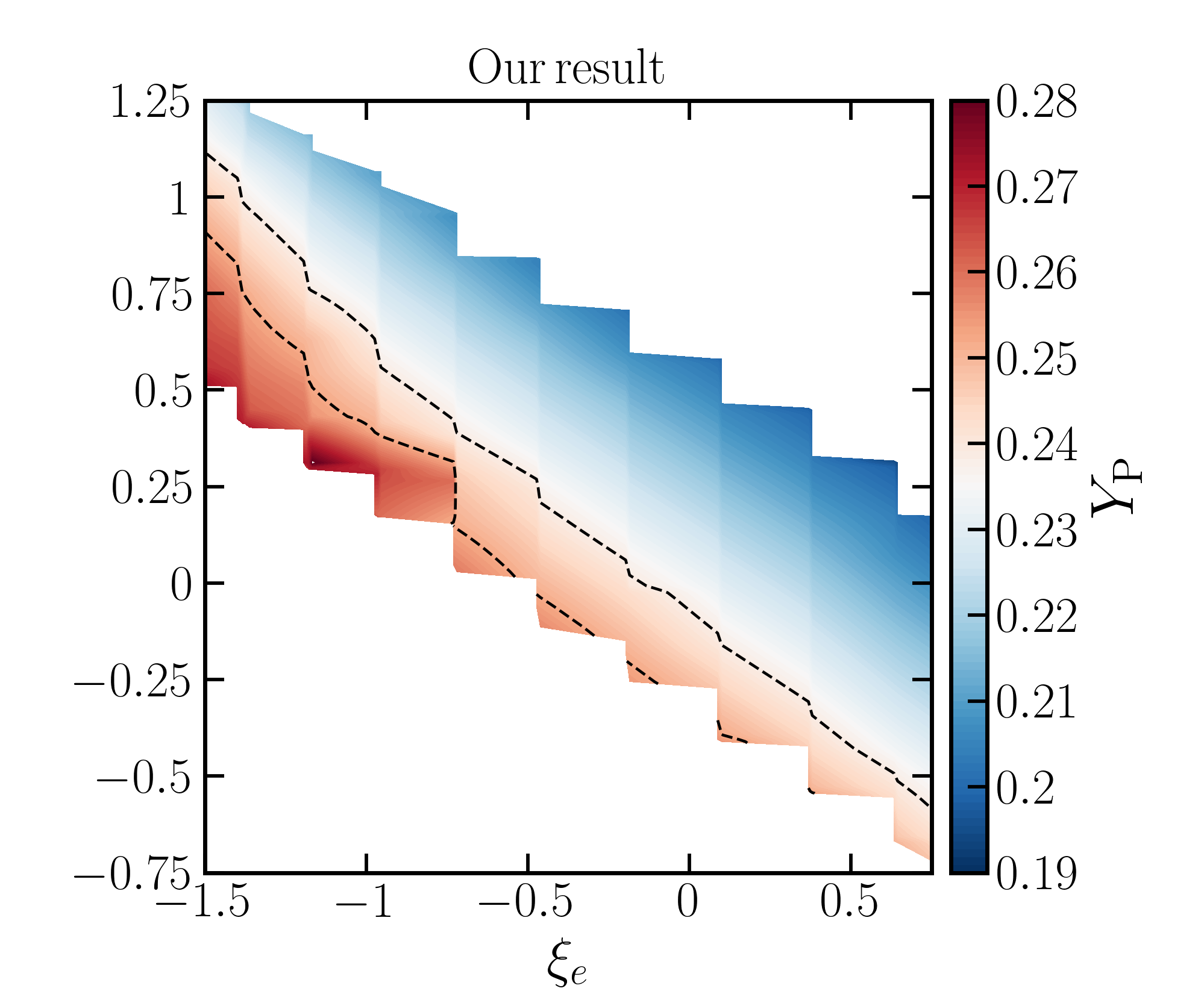}
\\
\hspace{-1.66cm} \includegraphics[width=0.44\textwidth]{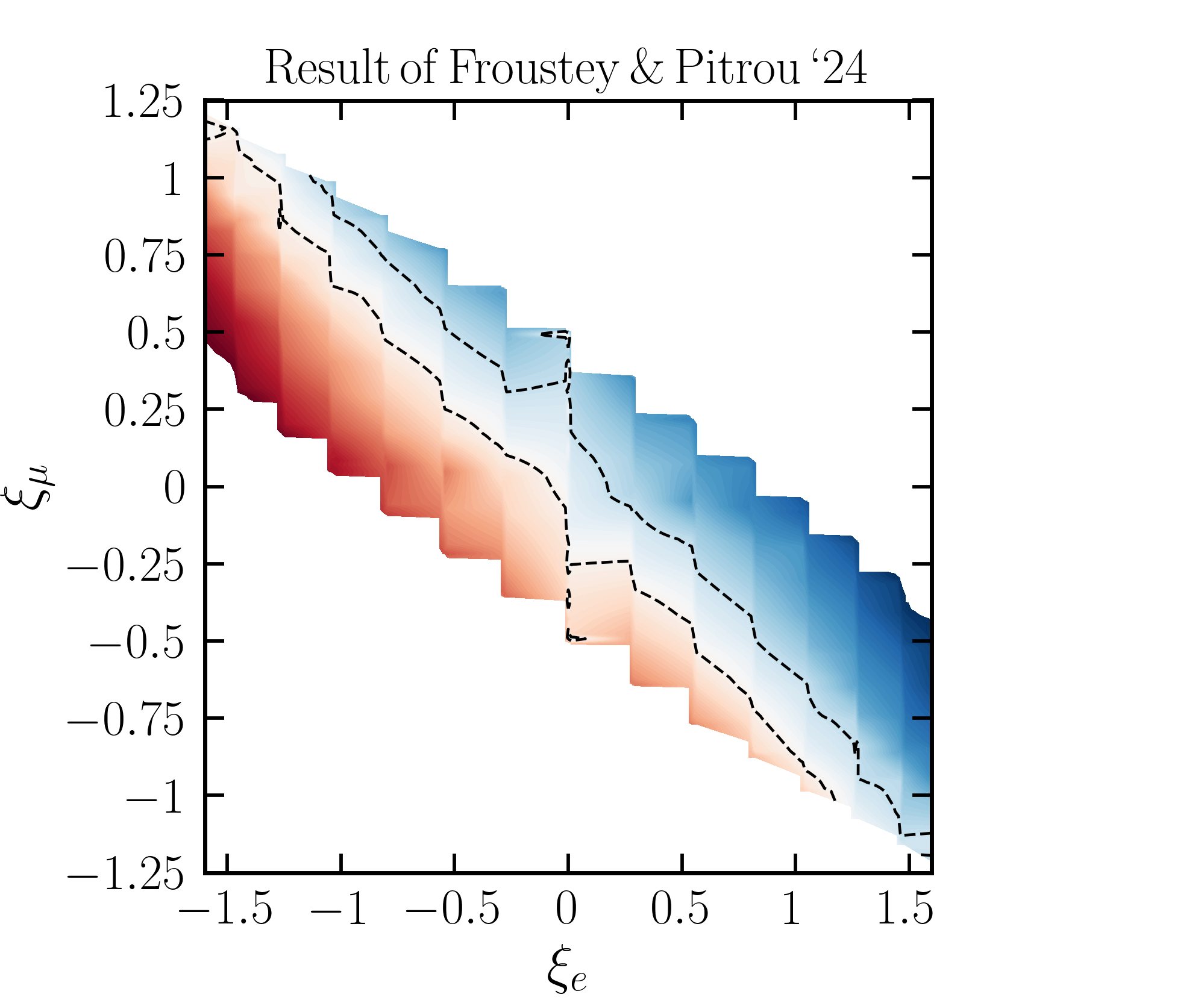} 
& \hspace{-1.75cm}  \includegraphics[width=0.44\textwidth]{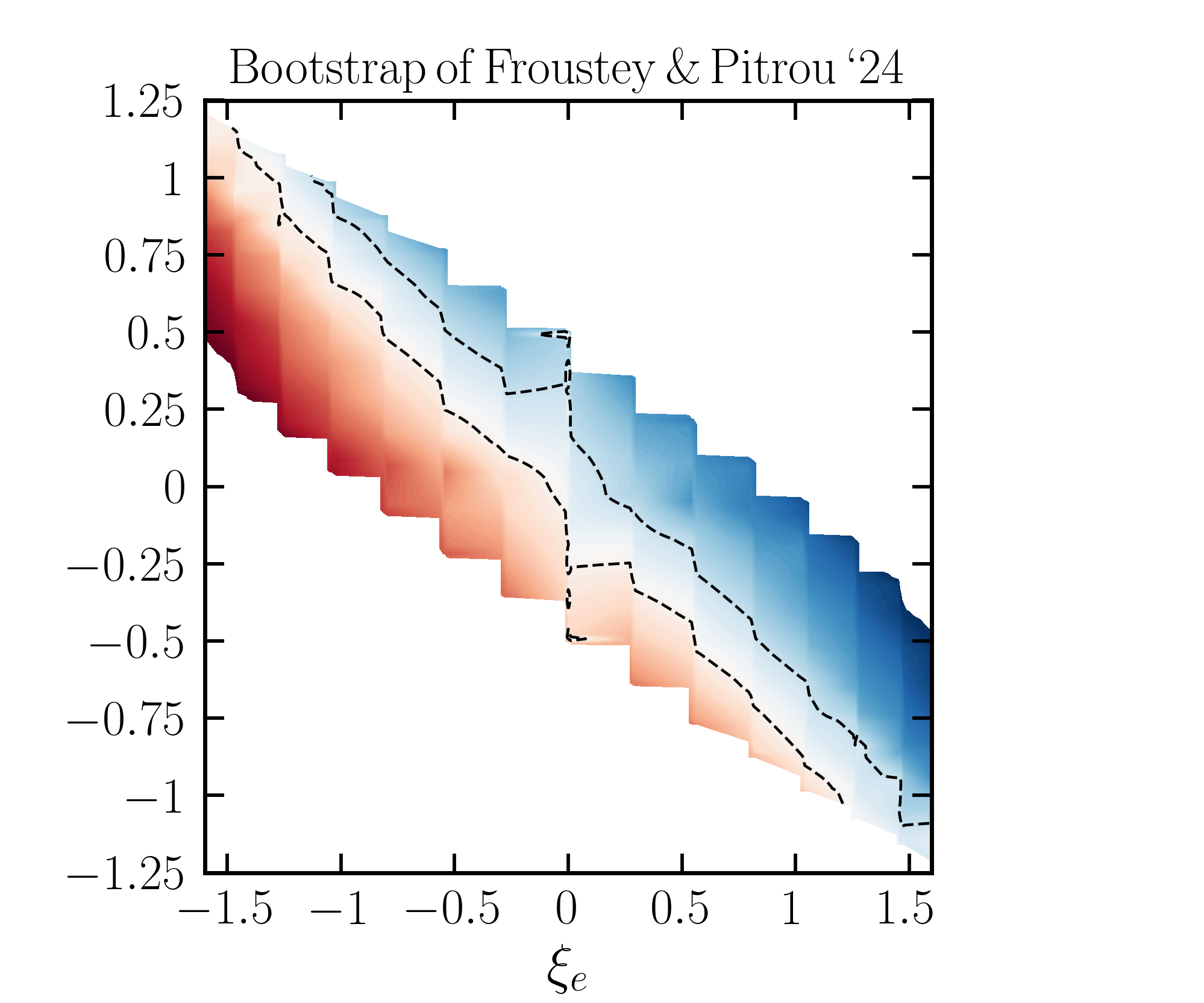}
& \hspace{-1.75cm}  \includegraphics[width=0.44\textwidth]{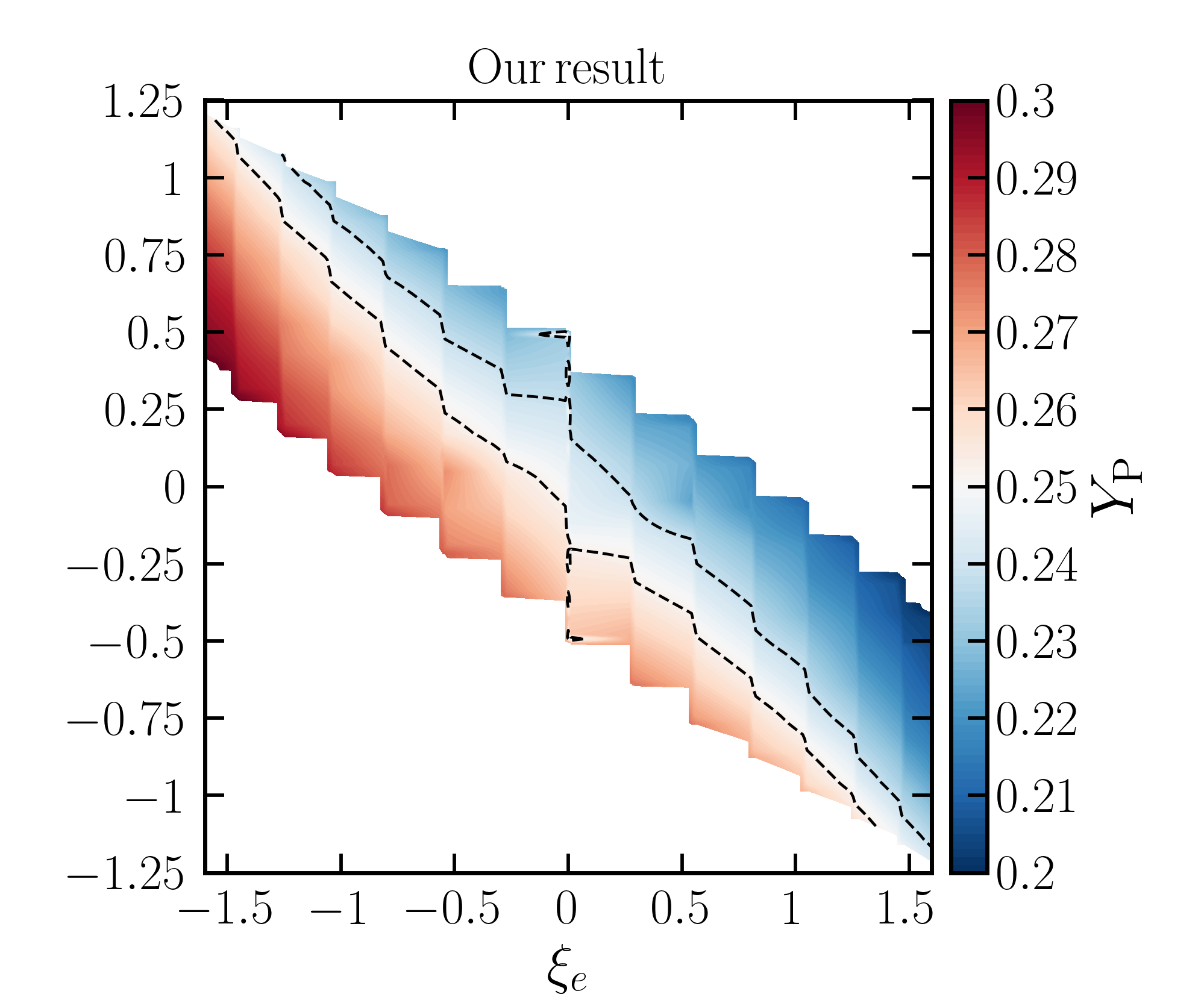}
\\
\hspace{-1.66cm} \includegraphics[width=0.44\textwidth]{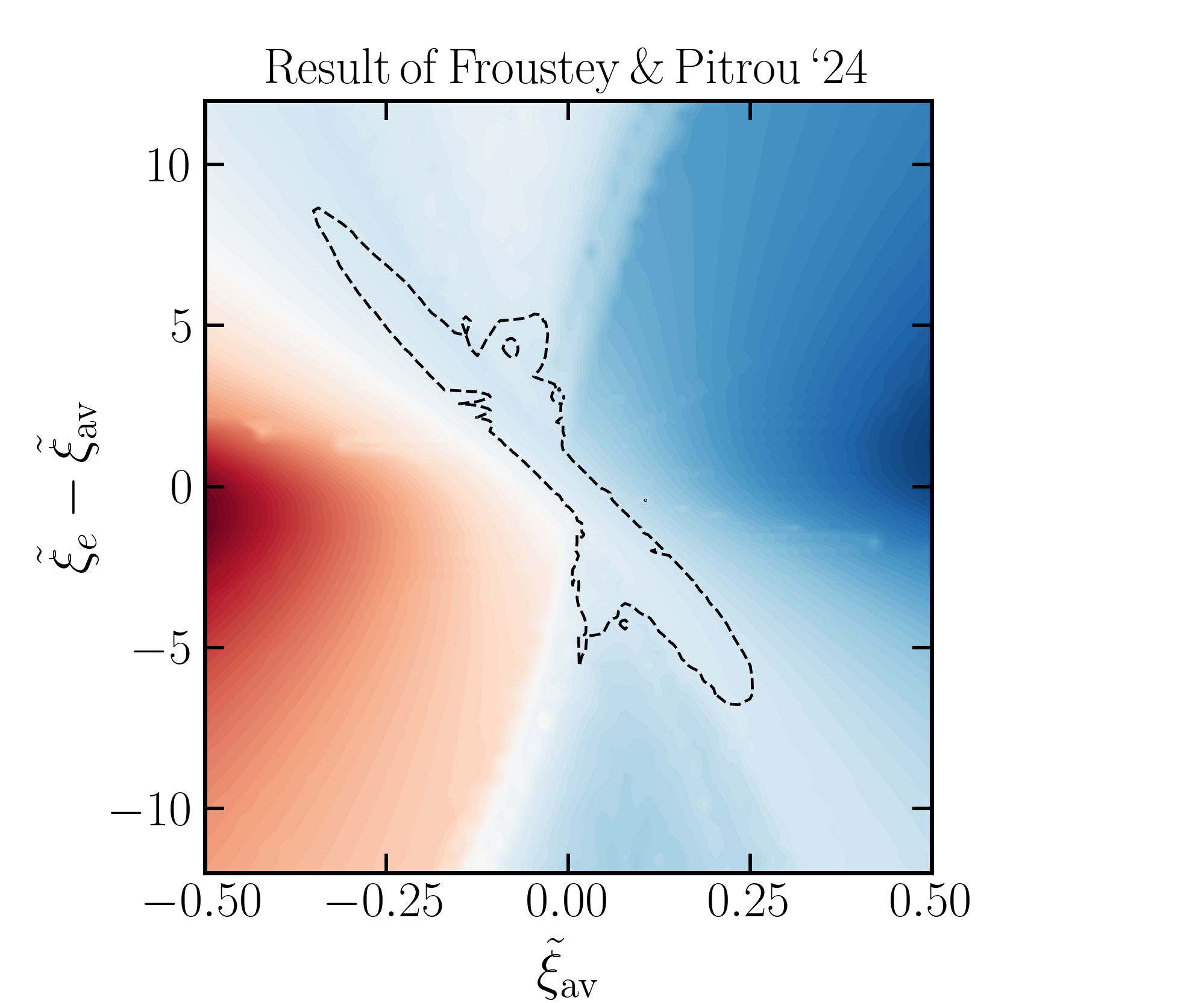} 
& \hspace{-1.75cm}  \includegraphics[width=0.44\textwidth]{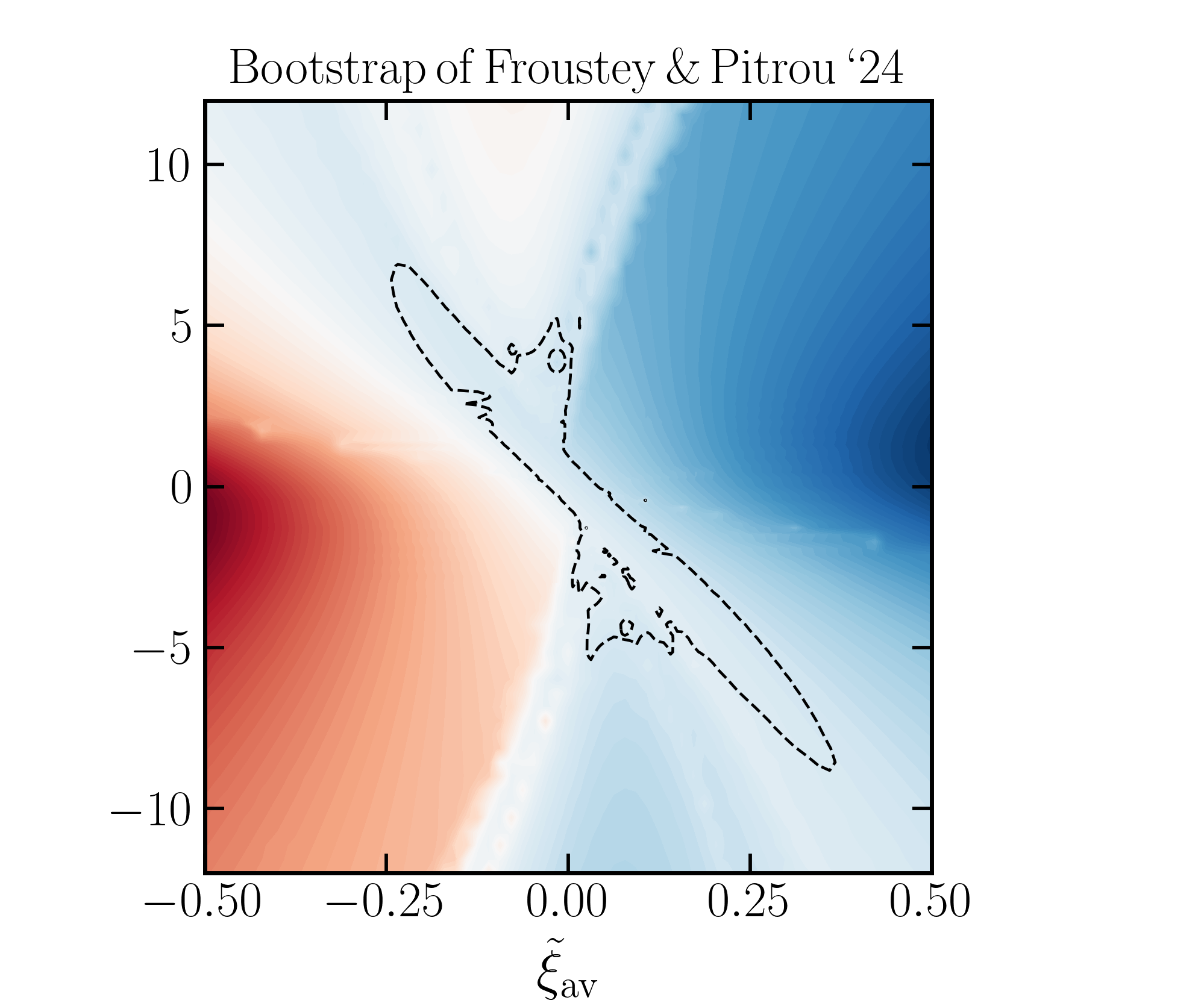}
& \hspace{-1.75cm}  \includegraphics[width=0.44\textwidth]{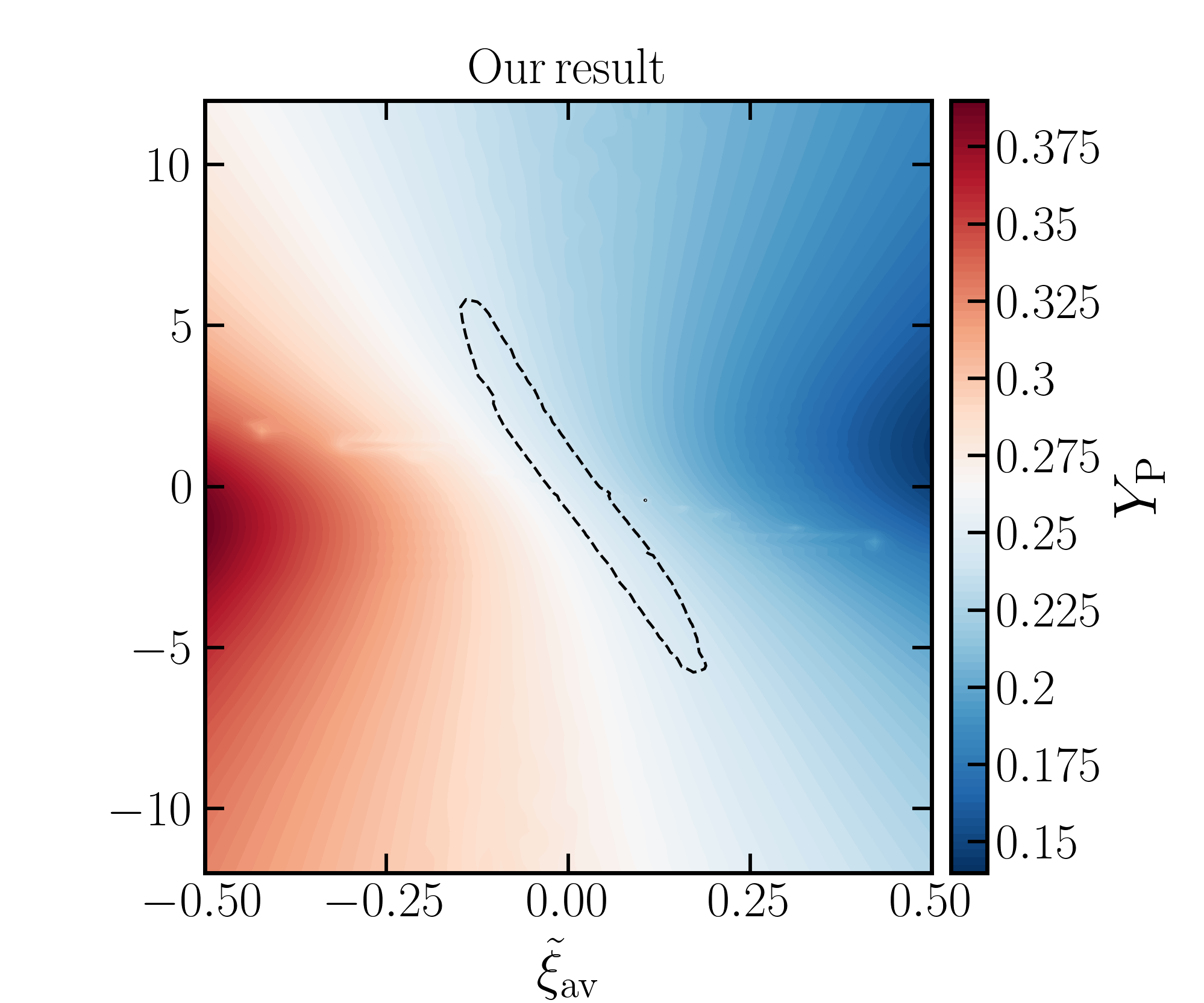}
\end{tabular}
\caption{
\textit{Left:} Results from~\cite{Froustey:2024mgf, froustey_2024_11185598}. \textit{Middle:} Results using the fitting functions of Eq.~\eqref{eq:fittingfuncs} but using as an input the asymptotic results for $({\rho_{\nu_e} + \rho_{{\bar \nu}_e}})/{T_{\rm cm}^4}$, $({n_{\nu_e} - n_{{\bar \nu}_e}})/{T_{\rm cm}^3}$, $z_\gamma$, and $N_{\rm eff}$ from~\cite{Froustey:2024mgf, froustey_2024_11185598}. In the last row we adopt $\tilde\xi_\alpha \equiv \xi_\alpha + \xi_\alpha^3/\pi^2$ and $\tilde \xi_{\rm av} \equiv (\tilde\xi_e+\tilde\xi_\mu + \tilde\xi_\tau)/3$ as defined in Ref.~\cite{Froustey:2024mgf}. \textit{Right:} Our results obtained with our momentum-averaged approach and our BBN pipeline. The dashed lines show the 95\% CL contours from our joint BBN+CMB analysis (see Eqs.~\eqref{eq:Ypmeasurement}-\eqref{eq:NeffPlanck}) and one can see an excellent agreement between all three panels. The first and second row correspond to regions of parameter space with $0.35 \geq \xi_e + \xi_\mu + \xi_\tau \geq 0.25$ and $0.05 \geq \xi_e + \xi_\mu + \xi_\tau \geq -0.05$ (i.e. approximately vanishing lepton number), respectively. The bottom row corresponds to $\xi_\mu = \xi_\tau$, for which, as identified in App.~\ref{sec:nonadiabatic}, a non-adiabatic second electron-driven MSW transition can occur at $T\lesssim 3\,\mathrm{MeV}$. In this special case we observe a mild disagreement between the left column and both the middle and right column. We remark that this is a very special region of parameter space, which can be considered a stress test of our momentum averaged approach.
}
\label{fig:xieximu_Zen}
\end{figure*}

Fig.~\ref{fig:xieximu_Zen} shows a direct comparison between the results of Ref.~\cite{Froustey:2024mgf} and summarizes the validation of our approach. The left panels show explicitly the results from Ref.~\cite{Froustey:2024mgf} from their publicly available tables \texttt{NEVO\_PRIMAT\_grid\_3D.csv} (upper and middle) and \texttt{NEVO\_PRIMAT\_grid\_xiav\_xie.csv} (lower)~\cite{froustey_2024_11185598}. In the middle panel we show the helium abundance that one would obtain by using our derived fitting functions in Eq.~\eqref{eq:fittingfuncs} using the results from Ref.~\cite{Froustey:2024mgf, froustey_2024_11185598} for the asymptotic neutrino densities at $T\simeq 1.5\,{\rm MeV}$, see Eq.~\eqref{eq:MapEqs}. The comparison between the left and middle panels clearly show that our approach in Sec.~\ref{subsec:BBN_Analysis} of using a Fermi-Dirac distribution for neutrinos is very accurate in predicting the helium abundance. Finally, in the right panels we show the predictions for the helium abundance from our solutions for the neutrino QKEs which then we feed into our BBN pipeline. For the two upper panels we can clearly see that the agreement is excellent across the parameter space. The one for the lower panel is slightly worse. However, one can clearly see that the region where we do not agree is significantly disfavored by data and falls within the highly non-trivial region of parameter space described in App.~\ref{sec:nonadiabatic}. We note that this is a very special region of parameter space, which can be considered a stress test of our momentum-averaged approach. 

All in all given the comparison with~\cite{Froustey:2024mgf} we believe that our joint approach to solve the neutrino ensemble using averaged neutrino distribution functions together with considering an effective Fermi-Dirac distribution for neutrinos to model weak neutron-proton conversion rates is in overall excellent agreement with the full solution using momentum dependent neutrino density matrices. 

Finally, we note that while we do not use the primordial deuterium abundance as a constraint in our analyses, we also find an excellent agreement in the comparison between our results and those of Ref.~\cite{Froustey:2024mgf}.

\setlength{\parskip}{1pt}

\cleardoublepage

\bibliography{biblio}

\end{document}